\documentclass{article}
\usepackage[dvips]{graphicx}

\title{Double complex SUSY-transformations: deformations of real potentials and their spectral characteristics}

\author{Sergei~P.~Maydanyuk
\thanks{E-mail: maidan@kinr.kiev.ua} \\
\small\emph{Institute for Nuclear Research, National Academy of Sciences of Ukraine,} \\
\small\emph{47, prosp. Nauki, Kiev-28, 03680, Ukraine}}

\date{\small\today}

\begin{document}

\maketitle

\vspace{-7mm}
\begin{abstract}
In paper \emph{approach of double complex SUSY-transformations with not coincident complex energies of transformation} is developed, allowing to deform given real potential $V_{1}$ (with keeping of its shape real) with obtaining exact solutions. The explicit solutions of the deformation of shape of the potential, its wave function at arbitrary energy, not coincident with energies of transformation, wave functions at the energies of transformation are obtained, condition of keeping of continuity of the solutions and isospectral condition are determined.
Using a rectangular well of finite width with infinitely high walls as the starting $V_{1}$ with discrete energy spectrum, by the proposed approach new types of deformation of this potential \emph{with deformation of the energy spectrum as a whole} have been obtained. The new potential contains the rectangular well as own partial case (with simultaneous transformation of the shape of this new potential, energy spectrum, wave functions of all bound states, wave function at arbitrary energy into corresponding characteristics of the rectangular well at needed choice of parameters).
Using null potential as the starting $V_{1}$ with continuous energy spectrum, new form of reflectionless real potential has been constructed. This potential generalizes well-known reflectionless potential of the type $V_{\rm ref}(x) = A^{2}(1-2\,{\rm sech}^{2}{Ax})$, allowing:
to pull down tails of the potential $V_{\rm ref}$ in the asymptotic regions up to zero (with keeping of nonzero depth);
to pull down \emph{continuously} the depth of the hole;
to displace arbitrary along axis $x$ the hole with its passing through zero;
to create and to increase the second hole, transforming $V_{\rm ref}$ into double-well potential;
to control continuously and simply the asymmetry of the shape of such reflectionless potential.
Note relative simplicity of the found potential in a comparison with variety of the reflectionless shape invariant potentials.
\end{abstract}





\newpage
\section*{Designations
\label{sec.0}}

In paper we shall use the following designations.
\begin{itemize}

\item
The first potential $V_{1}$, on the basis of which we shall construct new potentials using SUSY-transformations, we shall name as \emph{the starting potential with number ``1''}.

\item
$\varphi_{i}^{(k)}$ and $\varphi_{i}^{(k,1)}$, $\varphi_{i}^{(k,2)}$ --- general and two partial solutions for wave function (WF or WFs in the plural case) of the potential $V_{k}$ with the number $k$, where instead of index $i$ at the bottom we shall use one index from the following:
{\small
\begin{itemize}

\item
$n$ --- the number of the level of discrete energy spectrum of the potential $V_{k}$ with number $k$, if this level does not coincide with \emph{energy of transformation} $\cal E$ (i.~e. the energy, at which the superpotential is defined);

\item
$m$ --- the number of the level of the discrete energy spectrum of the potential $V_{k}$ with the number $k$, if this level coincides with the energy of transformation $\cal E$;

\item
$w$ --- the index at energy, if it coincides with the energy of transformation $\cal E$ and does not coincide with any level of the energy spectrum of the starting potential $V_{k}$.
\end{itemize}}

\item
$\bar{\varphi_{i}}^{(k)}$ --- WF of \emph{discontinuous state}, which we shall denote by stroke above.

\item
$E_{i}^{(k)}$ --- the level with the number $i$ (corresponding to WF $\varphi_{i}^{(k)}$) of the energy spectrum for the potential $V_{k}$. We shall number the levels of the discrete spectrum so:
{\small
\begin{itemize}

\item
the lowest level, concerned with the ground bound state, by number \emph{``1''};

\item
the next levels, located higher and described excited bound states, by numbers \emph{``2''} and larger.
\end{itemize}}

\item
$\phi_{i}^{(k)}$ --- \emph{function of transformation} (\emph{function of factorization}), defined as partial solution of $\varphi_{i}^{(k)}$ (for the bound or unbound or discontinuous state).

\item
${\cal E}_{i}^{(k)}$ --- \emph{energy of transformation} (\emph{energy of factorization}), defined at potential $V_{k}$ with the number $k$, coinciding with $E_{i}^{(k)}$.

\item
$W_{i}^{(k)}$ --- superpotential with the energy of transformation ${\cal E}_{i}^{(k)}$ and function of  transformation $\phi_{i}^{(k)}$ concerning potential $V_{k}$.

\item
Operators $A_{i}$ and $A_{i}^{+}$ we define so:
\begin{equation}
\begin{array}{cc}
  A_{i} = \displaystyle\frac{d}{dx} + W_{i}(x), &
  A_{i}^{+} = -\displaystyle\frac{d}{dx} + W_{i}(x).
\end{array}
\label{eq.0.1}
\end{equation}

\end{itemize}

Sometime, in text upper or lower index at such notations will be omitted.

\newpage
\section{Introduction
\label{sec.introduction}}

At present time an essential progress has been achieved in study of properties of quantum systems. Here, a great importance is directed to development of methods for control (management) of spectral characteristics of these systems, realized by inverse problem approach \cite{Zakhariev.1990.PEPAN,Zakhariev.1994.PEPAN,Zakhariev.1999.PEPAN,Zakhariev.2002.PEPAN},
and by approach of supersymmetric quantum mechanics (SUSY QM)
\cite{Cooper.1995.PRPLC,Roy.1990.IJMPA,Witten.1981.NuclPhys,Gendenshtein.1983.JETPL,%
Sukumar.1985.JPAGB.L57,Andrianov.1984,Andrianov.1984.TMP}.

In the approach of SUSY QM, ways of construction of new \emph{isospectral potentials} (with obtaining of exact solutions for wave function and energy spectrum) on the basis of the given one have been studied the most deeply \cite{Pursey.1986.PRD}. \emph{Darboux transformations} of the first order give this naturally, allowing to come from the given potential with discrete energy spectrum to new one with the energy spectrum without the lowest level (or with its little displacement for some potentials) \cite{Andrianov.1984,%
Sukumar.1985.JPAGB.L57,Sukumar.1985.JPAGB.p2917,Sukumar.1985.JPAGB.p2937,%
Sukumar.1986.JPAGB.p2297,Sukumar.1987.JPAGB.2461,Sukumar.1988.JPAGB.L455}
(see also reviews about these transformations \cite{Rosu.1998}, about \emph{factorization method} \cite{Infeld.1951.RMP,Mielnik.2004} and historical paper \cite{Darboux.1882}).
Transition to the Darboux transformations of \emph{the second order} gives a possibility to delete not only the lowest but arbitrary level in the energy spectrum, with obtaining explicit form for the new potential (without divergencies) and its wave functions \cite{Bagrov.1997.PEPAN,Samsonov.quant-ph.9904009,Samsonov.2002.PRC,Samsonov.2002.PLA}
(see also \cite{Andrianov.1995.IJMPA,Andrianov.1995.PLA}).
In definition of function of factorization a transition from wave function of bound state to its general solution gives a possibility to select energy of factorization arbitrary. Such a way allows to analyze what happens with the potential if to change continuously a slope of wave function at selected coordinate at arbitrary selected level or to shift continuously this level in the spectrum. The constructed by such a way new potentials can be considered as the continuous deformations of the given potential.
Transition to the Darboux transformations of \emph{the higher order} \cite{Bagrov.1997.PEPAN,Bagrov.quant-ph/9804032} and also development of \emph{methods of nonlinear supersymmetry} \cite{Andrianov.2003.NuclPhys,Andrianov.2004.JPAGB} give else richer algorithms of construction of new potentials with obtaining exact solutions of their spectral characteristics.
Another way to construct more expansive types of transformations with obtaining new exactly solvable potentials can be reduced to introduction of more complicated representations of the superpotential (with keeping of the order of the differential transformations). Here, we note \emph{method of extended supersymmetry} \cite{Pashnev.1986.TMP,Berezovoi.1987.TMP,Berezovoi.1988.TMP,Berezovoj.1991.ZP},
\emph{matrix variants of transformations of polylinear (nonlinear) supersymmetry} \cite{Samsonov.2004.JPA},
methods of solution of \emph{system of coupled discrete Schr\"{o}dinger equations} \cite{Cannata.1993.JPA,Suzko.2002.PAN},
generalizations of the formalism of SUSY-transformations into \emph{3 and multi-dimensional spaces}
\cite{Andrianov.1984,Andrianov.1984.TMP,Humi.1988.JPA,Gonzalez-Lopez.1998.JGP}.
Else one way to construct the new exactly solvable potentials (which can be considered as independent) consists in the development of new types of the differential transformations and their deformations (with keeping of the starting potential and obtaining of new hamiltonians of the Schr\"{o}dinger type), where the standard differential operators (for example, the standard Darboux operators) appear as a partial case only. One can refer to the methods of \emph{nonlinear supersymmetry}, the methods of \emph{hidden supersymmetry (hidden SUSY)}, developed in the approaches
\cite{Plyushchay.2007.AP.v322,Plyushchay.2007.JPA,Plyushchay.2007.PLB,%
Plyushchay.2001.NPB,Plyushchay.2000.IJMPA,Plyushchay.1996.AP.v245}.
Note papers \cite{Andrianov.2007.NPA,Sokolov.2007.NPA} with a possible last and full version of the method, unified these approaches.

However, inclusion of more complicated representations for the superpotential leads to more complicated representations for the potentials, obtained by the approach of SUSY QM. So, use of the complex superpotential gives new varieties of the potentials, which turn out complex inevitably \cite{Baye.1996.NPA,Andrianov.1999.IJMP,Chabanov.2001.IP,Deb.2003.PLA}.
By use of the SUSY QM methods the exact solutions for new classes of potentials with PT-symmetry have been opened \cite{Bender.1998.PRL,Bender.1999.JP,Znojil.1999.PLA,Znojil.2000.MG,Znojil.2001.PLA,%
Znojil.2002.JPA,Znojil.2006.JPA.p.10247} and properties of this symmetry have been studied
\cite{Znojil.2000.PLB,Levai.2001.JPA,Levai.2002.JPA,Znojil.2004.JPA,%
Levai.2004.CzJP.v54n10,Levai.2004.CzJP.v54n1,Znojil.2005.JPA.p.5041,%
Znojil.2006.JPA.p.4047,Cannata.2007.AP}.
In research of new types of the potentials with real values, a difficulty in analysis of their bound states, in determination of divergencies which can cross out found new solution, increases essentially. How do SUSY-transformations with the complex superpotential or its more complicated representation expand a set of the early known real potentials, how difficult can bound states be calculated, how is the energy spectrum deformed? Answer on these questions will point to effectiveness of inclusion of the complex superpotential or its more complicated representations into the SUSY-transformations in search of new forms of deformations of the real potentials.

The given paper is directed to study of these questions. We restrict ourselves by consideration of one-dimensional real potentials only. Using an approach of double Darboux transformations with complex superpotentials and not coincident energies of transformations as the SUSY-transformations, we shall analyze how they allow to construct new potentials on the basis of given one. The paper is organized so. In sec.~\ref{sec.2} a variant of scheme of determination of partial and general solutions for the superpotential, assuming it and the function of factorization to be complex and the given potential --- real, is presented. In sec.~\ref{sec.3} a formalism of the double transformations with not coincident complex energies of transformations is developed. It turns out difficult to make conclusion about effectiveness of these transformations in construction of new forms of the real potentials in a general case. Therefore, for analysis as the starting potential we shall use rectangular well of finite width with infinitely high walls in sec.~\ref{sec.example.rectangular} as example of system with completely discrete energy spectrum, and null potential in sec.~\ref{sec.example.reflectionless} as example of system with completely continuous energy spectrum. In sec.~\ref{sec.conclusions} conclusions are presented.

\section{Superpotential as a solution of Riccati equation
\label{sec.2}}

To clarify, which the most general form an unknown superpotential has, if we know only the potential, concerning with this superpotential in the most prevailing formalism (for example, as in~\cite{Cooper.1995.PRPLC} p.~275--277, \cite{Sukumar.1985.JPAGB.p2917} p.~2922-2923, 2925--2927, \cite{Andrianov.1984} p.~19--20,
at an energy of factorization, located not higher then the lowest level of energy spectrum of the given potential), we inevitably come to a \emph{problem of a solution of Riccati equation}, where as unknown function the superpotential is used.

Let's consider a quantum system with one-dimensional real potential $V_{1}(x)$ and discrete energy spectrum completely. One can find spectral characteristics (energy spectrum, wave functions) of this system, solving the Schr\"{o}dinger equation:
\begin{equation}
\begin{array}{ccl}
  \hat{h}_{1} \varphi^{(1)}_{m}(x) & = &
    \biggl(-\displaystyle\frac{d^{2}}{dx^{2}} + V_{1}(x) \biggr)\,
      \varphi^{(1)}_{m}(x) =
      E^{(1)}_{m} \varphi^{(1)}_{m}(x),
\end{array}
\label{eq.2.1.1}
\end{equation}
where
$\hat{h}_{1}$ is hamiltonian of the system,
$\varphi^{(1)}_{m}(x)$ is wave function (WF) of a state with a number $m$,
$E_{m}^{(1)}$ is energy level, corresponding to the state with the number $m$ and having real value,
$E_{1}^{(1)}$ is the lowest level of the energy spectrum with the number $1$.
For the given potential $V_{1}$ we introduce a superpotential $W(x)$, defining it by such a condition (for example, according to \cite{Cooper.1995.PRPLC}, p.~287--289, if ${\cal E} = E_{1}^{(1)}$):
\begin{equation}
  V_{1}(x) = W^{2}(x) - \displaystyle\frac{d W(x)}{dx} + {\cal E},
\label{eq.2.1.2}
\end{equation}
where ${\cal E}$ is constant, which can have arbitrary value and we name it as \emph{energy of transformation (or energy of factorization)}. In this paper $W(x)$ and ${\cal E}$ are complex.
Let the potential $V_{1}(x)$ be \emph{exactly solvable}, i.~e. one can write WFs for all states and the energy spectrum of the system with such a potential in the explicit analytical form. Let's assume, that we know a form of the potential $V_{1}(x)$ and a general solution of wave function $\varphi^{(1)}(x)$ (before imposing boundary conditions) at arbitrary energy  $E^{(1)}$. Also we assume, that we do not know the superpotential $W(x)$ and we shall find it. Condition (\ref{eq.2.1.2}), defining the unknown $W(x)$, is \emph{the Riccati equation}. Let's find the superpotential, solving this equation.

At first, let's find a partial solution of $W(x)$. Fulfilling substitution of variable:
\begin{equation}
  x \to z = z(x) = u(x) + i \, v(x),
\label{eq.2.1.3}
\end{equation}
with such condition:
\begin{equation}
  f(x) = \displaystyle\frac{d z(x)}{dx},
\label{eq.2.1.4}
\end{equation}
where $u(x), v(x)$ are real functions and $f(x)$ is arbitrary else. Write:
\begin{equation}
\begin{array}{lll}
  \displaystyle\frac{d}{dx} =
    \displaystyle\frac{du}{dx} \displaystyle\frac{d}{du} +  \displaystyle\frac{dv}{dx} \displaystyle\frac{d}{dv} =
    u_{x} \displaystyle\frac{d}{du} +  v_{x} \displaystyle\frac{d}{dv}, &
\hspace{7mm}
  u_{x} = \displaystyle\frac{du}{dx}, &
  v_{x} = \displaystyle\frac{dv}{dx}.
\end{array}
\label{eq.2.1.5}
\end{equation}
With new variables the equation (\ref{eq.2.1.2}) has such a form:
\begin{equation}
  V_{1}(u) - {\cal E} =
  W^{2}(u) - \displaystyle\frac{d W(u)}{dx} =
  W^{2}(u) - u_{x} \displaystyle\frac{d W(u)}{du} - v_{x} \displaystyle\frac{d W(u)}{dv}.
\label{eq.2.1.6}
\end{equation}

Let's introduce new function $W_{R}(z)$:
\begin{equation}
\begin{array}{cc}
  W_{R}(z) = -\displaystyle\frac{1}{z+z_{0}}, &

  z_{0} =
  \left\{
    \begin{array}{cl}
      -u_{0} - i v_{0}, & \mbox{at } u \le 0,\; v \le 0; \\
      -u_{0} + i v_{0}, & \mbox{at } u \le 0,\; v \ge 0; \\
       u_{0} - i v_{0}, & \mbox{at } u \ge 0,\; v \le 0; \\
       u_{0} + i v_{0}, & \mbox{at } u \ge 0,\; v \ge 0;
    \end{array}
  \right.
\end{array}
\label{eq.2.1.7}
\end{equation}
where $u$, $v$ are real positive constants, $z_{0}$ is parameter, used for exclusion of divergence of $W_{R}(z)$ at zero $z=0$. We obtain:
\begin{equation}
\begin{array}{ccl}
\vspace{2mm}
  \displaystyle\frac{d W_{R}(z)}{du} & = &
  \displaystyle\frac{d}{du} \biggl( - \displaystyle\frac{1}{u+iv + z_{0}} \biggr) =
  \displaystyle\frac{1}{(u+iv + z_{0})^{2}} =
  \displaystyle\frac{1}{(z+z_{0})^{2}}, \\

\vspace{1mm}
  \displaystyle\frac{d W_{R}(z)}{dv} & = &
  \displaystyle\frac{d}{dv} \biggl( -\displaystyle\frac{1}{u+iv + z_{0}} \biggr) =
  \displaystyle\frac{i}{(u+iv + z_{0})^{2}} =
  \displaystyle\frac{i}{(z+z_{0})^{2}}, \\

  W_{R}^{2}(z) & = & \displaystyle\frac{1}{(z+z_{0})^{2}}.
\end{array}
\label{eq.2.1.8}
\end{equation}
One can write:
\begin{equation}
\begin{array}{l}
  \alpha \displaystyle\frac{d W_{R}}{du} +
  \beta \displaystyle\frac{d W_{R}}{dv} =
  (\alpha + i \beta) \: W_{R}^{2} =
  \displaystyle\frac{\alpha + i \beta}{(z+z_{0})^{2}},
\end{array}
\label{eq.2.1.9}
\end{equation}
where $\alpha$ and $\beta$ are unknown arbitrary constants.
%
%
Taking into account of this, rewrite equation (\ref{eq.2.1.6}) so:
\begin{equation}
  V_{1}(u) - {\cal E} =
  W^{2}(u) - u_{x} \displaystyle\frac{d W(u)}{du} - v_{x} \displaystyle\frac{d W(u)}{dv} -
  f^{2}
  \biggl\{
    (\alpha + i \beta) W_{R}^{2} -
    \alpha \displaystyle\frac{d W_{R}}{du} -
    \beta \displaystyle\frac{d W_{R}}{dv}
  \biggr\}.
\label{eq.2.1.10}
\end{equation}
Fixing $f(u)$ by condition:
\begin{equation}
  W^{2}(x) = (\alpha + i\, \beta) \; f^{2}(u) \: W_{R}^{2}(u),
\label{eq.2.1.11}
\end{equation}
from (\ref{eq.2.1.10}) we find:
\begin{equation}
\begin{array}{lcl}
  V_{1}(u) - {\cal E} & = &
  - u_{x} \displaystyle\frac{d W(u)}{du} - v_{x} \displaystyle\frac{d W(u)}{dv} +
  \alpha f^{2} \displaystyle\frac{d W_{R}}{du} +
  \beta f^{2} \displaystyle\frac{d W_{R}}{dv} = \\
  & = &
  - u_{x} \sqrt{\alpha + i\, \beta} \; \displaystyle\frac{d f W_{R}}{du} -
  v_{x} \sqrt{\alpha + i\, \beta} \: \displaystyle\frac{d f W_{R}}{dv} +
  \alpha f^{2} \displaystyle\frac{d W_{R}}{du} +
  \beta f^{2} \displaystyle\frac{d W_{R}}{dv}.
\end{array}
\label{eq.2.1.12}
\end{equation}

Now we use $\alpha$ and $\beta$ so:
\begin{equation}
\begin{array}{lcl}
  \alpha f = u_{x} \sqrt{\alpha + i \,\beta}, &
  \beta f  = v_{x} \sqrt{\alpha + i \,\beta}.
\end{array}
\label{eq.2.1.13}
\end{equation}
Then from (\ref{eq.2.1.12}) we obtain:
\begin{equation}
\begin{array}{lcl}
\vspace{2mm}
  V_{1}(x) - {\cal E} & = &
  - u_{x} W_{R} \sqrt{\alpha + i\, \beta} \; \displaystyle\frac{d f}{du} -
  v_{x} W_{R} \sqrt{\alpha + i\, \beta} \: \displaystyle\frac{d f}{dv} =
  - W_{R} \sqrt{\alpha + i\, \beta} \;
  \Bigl( u_{x} \displaystyle\frac{d f}{du} + v_{x} \displaystyle\frac{d f}{dv} \Bigr) = \\

\vspace{2mm}
  & = &
  - W_{R} \sqrt{\alpha + i\, \beta} \;
  \Bigl( \displaystyle\frac{d f}{du} \displaystyle\frac{du}{dx} +
         \displaystyle\frac{d f}{dv} \displaystyle\frac{dv}{dx} \Bigr) =
  - W_{R} \sqrt{\alpha + i\, \beta} \; \displaystyle\frac{df}{dx} =
  - W_{R} \sqrt{\alpha + i\, \beta} \; \displaystyle\frac{d^{2} z(x)}{dx^{2}}.
\end{array}
\label{eq.2.1.14}
\end{equation}
Taking into account explicit form of (\ref{eq.2.1.7}) for $W_{R}$, we find:
\begin{equation}
  \Bigl( V_{1}(x) - {\cal E} \Bigr) \,\bigl(z(x)+z_{0}\bigr) =
  \sqrt{\alpha + i\, \beta} \; \displaystyle\frac{d^{2} \bigl(z(x)+z_{0}\bigr)}{dx^{2}}.
\label{eq.2.1.15}
\end{equation}
One can fulfill the condition (\ref{eq.2.1.13}), using $\alpha$ and $\beta$ so:
\begin{equation}
  \alpha + i\, \beta = 1.
\label{eq.2.1.16}
\end{equation}
In this case, equation (\ref{eq.2.1.15}) transforms into the following:
\begin{equation}
  -\displaystyle\frac{d^{2} \bigl(z(x)+z_{0}\bigr)}{dx^{2}} +
  V_{1}(x) \:\bigl(z(x)+z_{0}\bigr) =
  {\cal E} \,\bigl(z(x)+z_{0}\bigr).
\label{eq.2.1.17}
\end{equation}
Now we have obtained the Schr\"{o}dinger equation with the given real potential $V_{1}(x)$, where $z(x)+z_{0}$ is a complex wave function (we denote it as $\varphi_{\cal E}^{(1)}(x)$) for selected complex value $\cal E$.
So, let's write the partial solution $\tilde{W}(x)$ for the superpotential:
\begin{equation}
\begin{array}{ccl}
  z(x) & = & C_{0} \varphi^{(1)}_{\cal E}(x) - z_{0}, \\
  f(x) & = &
    \displaystyle\frac{dz(x)}{dx} =
    C_{0} \displaystyle\frac{d \varphi^{(1)}_{\cal E}(x)}{dx}, \\
  W_{R}(x) & = &
    -\displaystyle\frac{1}{z(x)+z_{0}} =
    -\displaystyle\frac{1}{C_{0} \: \varphi^{(1)}_{\cal E}(x)}, \\
  \tilde{W}(x) & = & f(x) \:W_{R}(z(x)) =
    -\displaystyle\frac{1}{\varphi^{(1)}_{\cal E}(x)}
    \displaystyle\frac{d \varphi^{(1)}_{\cal E}(x)}{dx}, \\
  C_{0} & = & const.
\end{array}
\label{eq.2.1.18}
\end{equation}
Further in this paper, we shall name function $\varphi^{(1)}_{\cal E}(x)$ (its two partial solutions $\varphi^{(1,1)}_{\cal E}(x)$ and $\varphi^{(1,2)}_{\cal E}(x)$), defining superpotential by such a way, as \emph{function of transformation} (according to logics in \cite{Bagrov.1997.PEPAN,Bagrov.quant-ph/9804032},
one can name it also as \emph{function of factorization}) and denote it $\phi(x)$ (the partial solutions as $\phi^{(1,1)}(x)$ and $\phi^{(1,2)}(x)$).

Analyzing logic of obtaining solution of Riccati equation and using the found partial solution $\tilde{W}$ for the superpotential, we find its general solution:
\begin{equation}
\begin{array}{cc}
  W(c_{new}) = \tilde{W} + \Delta W (c_{new}), &
  \Delta W(c_{new},x) = \displaystyle\frac{\mathstrut
    e^{\int \tilde{W}(x) \: dx}} {c_{new} - \displaystyle\int  e^{\int \tilde{W}(x) \: dx} dx},
\end{array}
\label{eq.2.1.19}
\end{equation}
where $c_{new}$ is new complex constant of integration.

\vspace{3mm}
If we know the partial solution for wave function $\varphi^{(1)}_{\cal E}(x)$ at any complex energy $\cal E$, then (\ref{eq.2.1.18}) and (\ref{eq.2.1.19}) determine the partial and general solutions for superpotential $W(x)$ at complex energy of transformation $\cal E$. A possibility of the energy of transformation to be complex introduces one new degree of freedom in definition of the superpotential $W(x)$ on the basis of (\ref{eq.2.1.2}), keeping the given real potential $V_{1}$. Therefore, a set of all functions of transformation $\varphi_{\cal E}(x)$ is essentially larger then a set of all eigenfunctions $\varphi^{(1)}_{n}(x)$, defined concerning the real potential $V_{1}$ and the discrete energy spectrum $E_{n}$.

In construction of new potentials-partners on the basis of the general solution $W(c_{new})$ the constant $c_{new}$ allows to deform their shapes. Another possibility to deform the new potential the imaginary part of $\cal E$ gives. Therefore, $c_{new}$ and $\cal E$ can be considered as \underline{independent} parameters of deformation of the new potentials.

\vspace{3mm}
\noindent
\underline{\bf Analysis:}

\begin{itemize}
\item
The solutions for superpotential are obtained at arbitrariness in a choice of boundary conditions, which can be imposed on the function of transformation. Therefore, definition of superpotential only on the basis of bound states is partial one.

\vspace{-1mm}
\item
The function of transformation can be not only WF of the bound state.

\vspace{-1mm}
\item
The energy of transformation $\cal{E}$ is arbitrary constant (it can be real, not coinciding with the ground or excited level $E^{(1)}_{m}$, or complex).

\vspace{-1mm}
\item
Solution (\ref{eq.2.1.13}) proves a possibility to construct the continuous potential $V_{1}$ with own bound states and without divergencies on the basis of discontinuous superpotential $W(x)$ with divergencies. Therefore, investigations of complex superpotentials with divergencies have a sense.

\vspace{-1mm}
\item
The found general solution (\ref{eq.2.1.19}) for superpotential, defined concerning one given real $V_{1}$, has two additional independent free parameters of deformation --- imaginary parts of $c_{new}$ and $\cal E$ (in a comparison with real superpotential for the same $V_{1}$).
\end{itemize}

\section{Double SUSY-transformations with not coincident energies of transformation: ${\cal E}^{(1)} \ne {\cal E}^{(2)}$
\label{sec.3}}

\subsection{Deformation of wave function at arbitrary energy $E_{k}$, not coincident with the energies of transformation: $E_{k} \ne {\cal E}^{(1)}$ and $E_{k} \ne {\cal E}^{(2)}$
\label{sec.3.1}}

Let's we know one partial solution $\varphi_{k}^{(1,1)}(x)$ for wave function at energy $E^{(1)}_{k}$ (not obviously coincident with the levels $E^{(1)}_{n}$ of the discrete energy spectrum) of the potential $V_{1}$. On its basis one can find the second partial solution $\varphi_{k}^{(1,2)}(x)$ for wave function at this energy so:
\begin{equation}
  \varphi_{k}^{(1,2)}(x) =
    \varphi_{k}^{(1,1)}(x)
    \int\limits_{x_{0}}^{x}
    \displaystyle\frac{dx'} {(\varphi_{k}^{(1,1)}(x'))^2},
\label{eq.3.1.1}
\end{equation}
and we construct the general solution:
\begin{equation}
  \varphi_{k}^{(1)}(c_{k}, x) =
    \varphi_{k}^{(1,1)}(x) + c_{k} \cdot \varphi_{k}^{(1,2)}(x) =
    \varphi_{k}^{(1,1)}(x) \:
    \biggl(1 + c_{k} \cdot\int\limits_{x_{0}}^{x}
    \displaystyle\frac{dx'} {(\varphi_{k}^{(1,1)}(x'))^2} \biggr),
\label{eq.3.1.2}
\end{equation}
where $c_{k}$ is arbitrary constant, which introduces one degree of freedom in deformation of the total wave function $\varphi_{k}^{(1)}$ (we add index os state $n$ to $c$). The value of $c_{k}$ one can fix by use of boundary conditions for $\varphi_{k}^{(1)}$ (we use this formalism from \cite{Berezovoi.1987.TMP}, p.~103--104; see also \cite{Berezovoj.1991.ZP}, p.~526). For arbitrary level $E_{n}^{(1)}$ sometimes it is convenient as the first solution  $\varphi_{n}^{(1,1)}$ to use \emph{WF of the bound state}. Then $c_{n}$ can be considered as \emph{parameter of deformation}, varying of which, we transform this state into \emph{unbound} one at level $E_{n}^{(1)}$, deforming the shape of $\varphi^{(1)}_{n}$.

In the first transition $V_{1} \to V_{2}$ we find a general solution of WF for $V_{2}$ at the arbitrary energy $E_{k}$ (where $E_{k} \ne {\cal E}^{(1)}$ and $E_{k} \ne {\cal E}^{(2)}$) so:
\begin{equation}
\begin{array}{ccc}
  \varphi_{k}^{(2)} = N_{k}^{(1)} A_{w_{1}}(c_{w_{1}}) \varphi_{k}^{(1)}(c_{k}), &
  A_{w_{1}}(c_{w_{1}}) = \displaystyle\frac{d}{dx} + W^{(1)}(c_{w_{1}}), &
  W^{(1)}(c_{w_{1}}) = -\displaystyle\frac{d}{dx} \ln{ \phi_{w_{1}}^{(1)}(c_{w_{1}}) },
\end{array}
\label{eq.3.1.3}
\end{equation}
where $\varphi_{k}^{(1)}$ is the general solution of WF with parameter $c_{k}$, $\phi_{w_{1}}^{(1)}$ is the first function of transformation at energy ${\cal E}^{(1)}$ with parameter $c_{w_{1}}$.
In contrast with construction of formalism for real Darboux transformations, we use in formula (\ref{eq.3.1.3}) a logarithm with complex argument. Using the formula for the logarithm with the complex argument:
\begin{equation}
\begin{array}{ccl}
  \log{z} = \log |z| + i \arg{z},
\end{array}
\label{eq.3.1.4}
\end{equation}
we rewrite the superpotential in (\ref{eq.3.1.3}) so:
\begin{equation}
\begin{array}{ccl}
  W^{(1)}(c_{w_{1}}) =
    - \displaystyle\frac{d}{dx}  \ln {\Bigl| \phi_{w_{1}}^{(1)}(c_{w_{1}}) \Bigr|} -
    i\,\displaystyle\frac{d}{dx} \arg {\Bigl( \phi_{w_{1}}^{(1)}(c_{w_{1}}) \Bigr)}.
\end{array}
\label{eq.3.1.5}
\end{equation}
One can see that \emph{complex definition of the superpotential is possible and it expands essentially possibilities in search of new types of real potentials} (in a comparison with using only real superpotentials). This formula can be useful in separation of imaginary part from real one in the new potentials.

In inverse transition $V_{2} \to V_{1}^{\rm (def)}$, we find the general solution for WF for $V_{1}^{\rm (def)}$ at the same energy $E_{k}$ so:
\begin{equation}
\begin{array}{ccc}
  \varphi_{k}^{(1,\:{\rm def})} = N_{k}^{(2)} A^{+}_{w_{2}}(c_{w_{2}}) \varphi_{k}^{(2)}(c_{k}), &
  A^{+}_{w_{2}}(c_{w_{2}}) = -\displaystyle\frac{d}{dx} + W^{(2)}(c_{w_{2}}), &
  W^{(2)}(c_{w_{2}}) = \displaystyle\frac{d}{dx} \ln{ \phi_{w_{2}}^{(2)}(c_{w_{2}}) },
\end{array}
\label{eq.3.1.6}
\end{equation}
where $\phi_{w_{2}}^{(2)}$ is the second function of transformation at energy ${\cal E}^{(2)}$ with parameter $c_{w_{2}}$.

From (\ref{eq.3.1.3}) and (\ref{eq.3.1.6}) we obtain WF at the energy $E_{k}$ for the deformed $V_{1}^{\rm (def)}$:
\begin{equation}
  \varphi_{k}^{(1,\:{\rm def})} (c_{k},c_{w_{1}},c_{w_{2}}) =
    N_{k}^{(1)} N_{k}^{(2)} \bigl( E_{k} - {\cal E}^{(1)} \bigr)\, \varphi_{k}^{(1)} +
    N_{k}^{(1)} N_{k}^{(2)} \Delta W (c_{w_{1}},c_{w_{2}}) A_{w_{1}} \; \varphi_{k}^{(1)},
\label{eq.3.1.7}
\end{equation}
where $\Delta W (c_{w_{1}},c_{w_{2}}) = W^{(2)}(c_{w_{2}}) - W^{(1)}(c_{w_{1}})$.
We find $\Delta W (c_{w_{1}},c_{w_{2}})$:
\begin{equation}
  \Delta W (w_{1},w_{2}) =
    \displaystyle\frac
      {\varphi_{w_{1}}^{(1)}
        \Bigl(\varphi_{w_{2}}^{(1)}\Bigr)^{\prime\prime} -
        \Bigl(\varphi_{w_{1}}^{(1)} \Bigr)^{\prime\prime}
        \varphi_{w_{2}}^{(1)}}
      {\varphi_{w_{1}}^{(1)} \Bigl(\varphi_{w_{2}}^{(1)}\Bigr)^{\prime} -
        \Bigl(\varphi_{w_{1}}^{(1)} \Bigr)^{\prime} \varphi_{w_{2}}^{(1)}}.
\label{eq.3.1.8}
\end{equation}
Taking into account:
\begin{equation}
\begin{array}{cc}
  -\Bigl(\varphi_{w_{1}}^{(1)}\Bigr)^{\prime\prime} =
    ({\cal E}^{(1)} - V_{1}) \varphi_{w_{1}}^{(1)}, &
  -\Bigl(\varphi_{w_{2}}^{(1)}\Bigr)^{\prime\prime} =
    ({\cal E}^{(2)} - V_{1}) \varphi_{w_{2}}^{(1)},
\end{array}
\label{eq.3.1.9}
\end{equation}
we obtain:
\begin{equation}
  \Delta W (w_{1},w_{2}) =
    \displaystyle\frac
      { \Bigl( {\cal E}^{(1)} - {\cal E}^{(2)} \Bigr)
        \varphi_{w_{1}}^{(1)} \varphi_{w_{2}}^{(1)}}
      {\varphi_{w_{1}}^{(1)} \Bigl(\varphi_{w_{2}}^{(1)}\Bigr)^{\prime} -
        \Bigl(\varphi_{w_{1}}^{(1)} \Bigr)^{\prime} \varphi_{w_{2}}^{(1)}}.
\label{eq.3.1.10}
\end{equation}
Using this, from (\ref{eq.3.1.7}) we find the final form of WF at the energy $E_{k}$:
\begin{equation}
  \varphi_{k}^{(1,\:{\rm def})} (c_{k},c_{w_{1}},c_{w_{2}}) =
    N_{k}^{(1)} N_{k}^{(2)} \bigl( E_{k} - {\cal E}^{(1)} \bigr)\, \varphi_{k}^{(1)} +
    N_{k}^{(1)} N_{k}^{(2)}
      \Bigl( {\cal E}^{(1)} - {\cal E}^{(2)} \Bigr)  \varphi_{w_{2}}^{(1)}
    \displaystyle\frac
      {\varphi_{w_{1}}^{(1)} \Bigl(\varphi_{k}^{(1)}\Bigr)^{\prime} -
        \Bigl(\varphi_{w_{1}}^{(1)} \Bigr)^{\prime} \varphi_{k}^{(1)}}
      {\varphi_{w_{1}}^{(1)} \Bigl(\varphi_{w_{2}}^{(1)}\Bigr)^{\prime} -
        \Bigl(\varphi_{w_{1}}^{(1)} \Bigr)^{\prime} \varphi_{w_{2}}^{(1)}}.
\label{eq.3.1.11}
\end{equation}

If as the first or the second function of transformation, WF of the bound state at level ${\cal E}^{(1)}$ or ${\cal E}^{(2)}$ is used, then one can normalize this function of transformation and obtain normalizing $N_{k}^{(1)}$ and $N_{k}^{(2)}$ from this:
\begin{equation}
\begin{array}{cc}
  N_{k}^{(1)} =
  \left \{
  \begin{array}{cl}
    (E_{k} - {\cal E}^{(1)})^{-1/2},  & \mbox{at } E_{k} > {\cal E}^{(1)}, \\
    i|E_{k} - {\cal E}^{(1)}|^{-1/2}, & \mbox{at } E_{k} < {\cal E}^{(1)};
  \end{array}
  \right.
  &
  N_{k}^{(2)} =
  \left \{
  \begin{array}{cl}
    (E_{k} - {\cal E}^{(2)})^{-1/2},  & \mbox{at } E_{k} > {\cal E}^{(2)}, \\
    i|E_{k} - {\cal E}^{(2)}|^{-1/2}, & \mbox{at } E_{k} < {\cal E}^{(2)}.
  \end{array}
  \right.
\end{array}
\label{eq.3.1.12}
\end{equation}

\vspace{5mm}
\noindent
\underline{\bf Analysis:}

\vspace{-1mm}
\begin{itemize}

\item
Let's formulate a \emph{sufficient condition} from (\ref{eq.3.1.11}), when WF of the arbitrary state at arbitrary energy $E_{k}$, not coincident with both energies of transformation ${\cal E}^{(1)}$ and ${\cal E}^{(2)}$, remains continuous after deformation:
\begin{equation}
\begin{array}{ccl}
  \phi_{w_{1}}^{(1)}(x) \Bigl(\phi_{w_{2}}^{(1)}(x)\Bigr)^{\prime} -
  \Bigl(\phi_{w_{1}}^{(1)}(x) \Bigr)^{\prime} \phi_{w_{2}}^{(1)}(x)
  \ne 0.
\end{array}
\label{eq.3.1.13}
\end{equation}
This condition agrees with the requirement of inequality of Wronskian of two transformation functions to zero to provide the absence of singularities in the potential $V_{2}$, obtained on the basis of the given potential $V_{0}$ by use of Darboux transformations of the second order in~\cite{Samsonov.quant-ph.9904009}.

\vspace{-1mm}
\item
Fulfillment of condition (\ref{eq.3.1.13}) provides continuity of WFs of all states at arbitrary levels $E_{k}$ simultaneously, which are continuous for the starting $V_{1}$.

\vspace{-1mm}
\item
According to (\ref{eq.3.1.11}) and (\ref{eq.3.1.13}), all bound states of the starting potential $V_{1}$ remain bound after the deformation (with possible exclusion of states at the energies ${\cal E}^{(1)}$ and ${\cal E}^{(2)}$) at fulfillment of one from 4 conditions:
\begin{equation}
\begin{array}{cll}
  1) & \phi_{w_{1}}^{(1)} (x_{0}) = 0, &
       \phi_{w_{1}}^{(1)} (x_{1}) = 0; \\

  2) & \phi_{w_{2}}^{(1)} (x_{0}) = 0, &
       \phi_{w_{2}}^{(1)} (x_{1}) = 0; \\

  3) & \phi_{w_{1}}^{(1)} (x_{0}) = 0, &
       \phi_{w_{2}}^{(1)} (x_{1}) = 0; \\

  4) & \phi_{w_{2}}^{(1)} (x_{0}) = 0, &
       \phi_{w_{1}}^{(1)} (x_{1}) = 0.
\end{array}
\label{eq.3.1.14}
\end{equation}

\vspace{-1mm}
\item
In particularly, all bound states of the starting potential $V_{1}$ remain bound after the deformation, if the energy of transformation ${\cal E}^{(1)}$ or ${\cal E}^{(2)}$ coincides with arbitrary level of the spectrum for $V_{1}$, and as the function of transformation the WF of the bound state at this level is used.

\vspace{-1mm}
\item
All bound states of the starting $V_{1}$ remain bound after the deformation (with possible exclusion of states at energies ${\cal E}^{(1)}$ ш ${\cal E}^{(2)}$) \emph{at arbitrary energies ${\cal E}^{(1)}$ and ${\cal E}^{(2)}$}, if one from conditions is fulfilled:
\begin{equation}
\begin{array}{cll}
  1) & \phi_{w_{1}}^{(1)} (x_{0}) = 0, &
       \phi_{w_{2}}^{(1)} (x_{1}) = 0; \\

  2) & \phi_{w_{1}}^{(1)} (x_{1}) = 0, &
       \phi_{w_{2}}^{(1)} (x_{0}) = 0.
\end{array}
\label{eq.3.1.15}
\end{equation}
\end{itemize}

\subsection{Deformation of the potential
\label{sec.3.2}}

The shape of the potential $V_{1}$ is deformed so:
\begin{equation}
\begin{array}{ccl}
  V_{1}^{\rm (def)} (c_{w_{1}},c_{w_{2}}) & = &
    V_{1} + \Delta V_{1}(c_{w_{1}},c_{w_{2}}), \\
  \Delta V_{1} (c_{w_{1}},c_{w_{2}}) & = &
    2 \Bigl( {\cal E}^{(2)} - {\cal E}^{(1)} \Bigr)
      \displaystyle\frac{d}{dx} \displaystyle\frac
      {\varphi_{w_{1}}^{(1)} \varphi_{w_{2}}^{(1)}}
      {\varphi_{w_{1}}^{(1)} \Bigl(\varphi_{w_{2}}^{(1)}\Bigr)^{\prime} -
        \Bigl(\varphi_{w_{1}}^{(1)} \Bigr)^{\prime} \varphi_{w_{2}}^{(1)}}.
\end{array}
\label{eq.3.2.1}
\end{equation}
To be the new potential $V_{1}^{\rm (def)}$ real, we introduce the following condition:
\begin{equation}
  \Im{\Bigl( \Delta V_{1} (c_{w_{1}},c_{w_{2}, x}) \Bigr)} = 0.
\label{eq.3.2.2}
\end{equation}

\noindent
\underline{\bf Analysis:}

\vspace{-1mm}
\begin{itemize}

\item
The potential $V_{1}^{\rm (def)}$ has 4 independent complex parameters of deformation: ${\cal E}_{1}$,  ${\cal E}_{2}$, $c_{w_{1}}$, $c_{w_{2}}$.

\item
Continuity of the potential $V_{1}$, continuity of all WFs at all levels and energies (with possible exclusion of ${\cal E}^{(1)}$ and ${\cal E}^{(2)}$) after the deformation is defined by condition (\ref{eq.3.1.13}).
\end{itemize}

\subsection{Deformation of wave function at the energy of transformation ${\cal E}^{(1)}$
\label{sec.3.3}}

Let's find how WF of arbitrary state at the energy of transformation ${\cal E}^{(1)}$ is deformed. According to (\ref{eq.3.1.3}) and (\ref{eq.3.1.6}), one can find the general solution of the deformed WF so:
\begin{equation}
  \varphi_{k_{1}}^{(1,\:{\rm def})} =
    N_{w_{1}}^{(2)} A^{+}_{w_{2}} \varphi_{k_{1}}^{(2)} =
    N_{w_{1}}^{(2)}
      \biggl( -\displaystyle\frac{d}{dx} + W^{(2)} \biggr)
      \varphi_{k_{1}}^{(2)} =
    N_{w_{1}}^{(2)}
      \displaystyle\frac{
          -\varphi_{k_{1}}^{(2),\prime} \phi_{w_{2}}^{(2)} +
          \varphi_{k_{1}}^{(2)} \phi_{w_{2}}^{(2),\prime} }
        {\phi_{w_{2}}^{(2)}}.
\label{eq.3.3.1}
\end{equation}
We write the general solution for $\phi_{w_{2}}^{(2)}$:
\begin{equation}
  \phi_{w_{2}}^{(2)} =
    N_{w_{2}}^{(1)} A_{w_{1}} \phi_{w_{2}}^{(1)} =
    N_{w_{2}}^{(1)}
      \biggl( \displaystyle\frac{d}{dx} + W^{(1)} \biggr)
      \phi_{w_{2}}^{(1)} =
    N_{w_{2}}^{(1)}
      \displaystyle\frac{
          \phi_{w_{1}}^{(1)} \phi_{w_{2}}^{(1),\prime} -
          \phi_{w_{1}}^{(1),\prime} \phi_{w_{2}}^{(1)} }
        {\phi_{w_{1}}^{(1)}}
\label{eq.3.3.2}
\end{equation}
and its derivative:
\begin{equation}
\begin{array}{ccl}
  \phi_{w_{2}}^{(2), \prime} & = &
   N_{w_{2}}^{(1)}
      \displaystyle\frac{
          \Bigl({\cal E}^{(1)} - {\cal E}^{(2)} \Bigr)
          \phi_{w_{1}}^{(1)}
          \phi_{w_{1}}^{(1)}
          \phi_{w_{2}}^{(1)} -
          \phi_{w_{1}}^{(1)}
          \phi_{w_{1}}^{(1),\prime}
          \phi_{w_{2}}^{(1),\prime} +
          \phi_{w_{1}}^{(1),\prime}
          \phi_{w_{1}}^{(1),\prime}
          \phi_{w_{2}}^{(1)} }
        {\Bigl(\phi_{w_{1}}^{(1)}\Bigr)^{2}}.
\end{array}
\label{eq.3.3.3}
\end{equation}

It needs to take into account, that the wave function $\varphi_{w_{1}}^{(1)}$ before its deformation can differ from the used function of transformation $\phi_{w_{1}}^{(1)}$ at energy ${\cal E}^{(1)}$. If such wave function coincides with the function of transformation $\phi_{w_{1}}^{(1)}$, then the general solution for $\varphi_{k_{1}}^{(2)}$
can be found so:
\begin{equation}
\begin{array}{cc}
  \varphi_{k_{1}}^{(2)} =
    \phi_{w_{1}}^{(2)} =
    \displaystyle\frac{I_{w_{1}}^{(1)} + \lambda_{1}}{\phi_{w_{1}}^{(1)}}, &
  I_{w_{1}}^{(1)} =
    \displaystyle\int\limits_{x_{0}}^{x} \Bigl(\phi_{w_{1}}^{(1)} (x')\Bigr)^{2} dx'.
\end{array}
\label{eq.3.3.4}
\end{equation}
Finding the derivative of this function:
\begin{equation}
  \varphi_{k_{1}}^{(2),\prime} =
  \biggl(\displaystyle\frac{I_{w_{1}}^{(1)} + \lambda_{1}}
          {\phi_{w_{1}}^{(1)}} \biggr)^{\prime} =
  \phi_{w_{1}}^{(1)} -
    \phi_{w_{1}}^{(1),\prime}
    \displaystyle\frac{I_{w_{1}}^{(1)} + \lambda_{1}}
          {\Bigl( \phi_{w_{1}}^{(1)} \Bigr)^{2}}
\label{eq.3.3.5}
\end{equation}
we obtain:
\begin{equation}
  \varphi_{w_{1}}^{(1,\:{\rm def})} =
    -N_{w_{1}}^{(2)} \phi_{w_{1}}^{(1)} +
    N_{w_{1}}^{(2)} \Bigl({\cal E}^{(1)} - {\cal E}^{(2)}\Bigr)
      \displaystyle\frac
        { \Bigl( I_{w_{1}}^{(1)} + \lambda_{1} \Bigr)
                 \phi_{w_{2}}^{(1)} }
        { \phi_{w_{1}}^{(1)} \phi_{w_{2}}^{(1),\prime} -
          \phi_{w_{1}}^{(1),\prime} \phi_{w_{2}}^{(1)} }.
\label{eq.3.3.6}
\end{equation}

\vspace{5mm}
If wave function $\varphi_{k_{1}}^{(1)}$ before its deformation does not coincide with the function of transformation $\phi_{w_{1}}^{(1)}$ at energy ${\cal E}^{(1)}$, then the function:
\begin{equation}
  \varphi_{k_{1}}^{(2,1)} = \displaystyle\frac{\lambda_{1}}{\varphi_{k_{1}}^{(1)}}
\label{eq.3.3.7}
\end{equation}
is not a solution of the Schr\"{o}dinger equation with the new potential $V_{2}$ at such energy.
But in such a case, \emph{non-zero} function $\varphi_{k_{1}}^{(2)}$ can be found so:
\begin{equation}
  \varphi_{k_{1}}^{(2)} =
    N_{w_{1}}^{(1)} A_{w_{1}} \varphi_{k_{1}}^{(1)} =
    N_{w_{1}}^{(1)}
      \biggl( \displaystyle\frac{d}{dx} + W^{(1)} \biggr)
      \varphi_{k_{1}}^{(1)} =
    N_{w_{1}}^{(1)}
      \displaystyle\frac
        {\phi_{w_{1}}^{(1)} \varphi_{k_{1}}^{(1),\prime} - \phi_{w_{1}}^{(1),\prime} \varphi_{k_{1}}^{(1)} }
        {\phi_{w_{1}}^{(1)}},
\label{eq.3.3.8}
\end{equation}
where $N_{w_{1}}^{(2)}$ is new normalizing factor. Calculating the derivative of this function:
\begin{equation}
\begin{array}{ccl}
  \varphi_{k_{1}}^{(2), \prime} & = &
   N_{w_{1}}^{(1)} \cdot
      \displaystyle\frac{
          \phi_{w_{1}}^{(1),\prime}
          \phi_{w_{1}}^{(1),\prime}
          \varphi_{k_{1}}^{(1)} -
          \phi_{w_{1}}^{(1)}
          \phi_{w_{1}}^{(1),\prime}
          \varphi_{k_{1}}^{(1),\prime}}
        {\Bigl(\phi_{w_{1}}^{(1)}\Bigr)^{2}}.
\end{array}
\label{eq.3.3.9}
\end{equation}
we find:
\begin{equation}
  \varphi_{k_{1}}^{(1,\:{\rm def})} =
    N_{w_{1}}^{(1)} N_{w_{1}}^{(2)} \:
    \Bigl({\cal E}^{(1)} - {\cal E}^{(2)} \Bigr) \:
    \displaystyle\frac
      {\phi_{w_{1}}^{(1)} \varphi_{k_{1}}^{(1),\prime} - \phi_{w_{1}}^{(1),\prime} \varphi_{k_{1}}^{(1)} }
      {\phi_{w_{1}}^{(1)} \phi_{w_{2}}^{(1),\prime} - \phi_{w_{1}}^{(1),\prime} \phi_{w_{2}}^{(1)} } \:
    \phi_{w_{2}}^{(1)}.
\label{eq.3.3.10}
\end{equation}
Here, one can separate the deformation of $\varphi_{k_{1}}^{(1)}$ explicitly:
\begin{equation}
  \varphi_{k_{1}}^{(1,\:{\rm def})} =
  N_{w_{1}}^{(1)} N_{w_{1}}^{(2)}\,
    \Bigl({\cal E}^{(1)} - {\cal E}^{(2)} \Bigr) \:
    \varphi_{k_{1}}^{(1)} +
  N_{w_{1}}^{(1)} N_{w_{1}}^{(2)} \,
    \Bigl({\cal E}^{(1)} - {\cal E}^{(2)} \Bigr) \:
    \displaystyle\frac
      {\phi_{w_{2}}^{(1)} \varphi_{k_{1}}^{(1),\prime} - \phi_{w_{2}}^{(1),\prime} \varphi_{k_{1}}^{(1)} }
      {\phi_{w_{1}}^{(1)} \phi_{w_{2}}^{(1),\prime} - \phi_{w_{1}}^{(1),\prime} \phi_{w_{2}}^{(1)} } \:
    \phi_{w_{1}}^{(1)}.
\label{eq.3.3.11}
\end{equation}

\subsection{Deformation of wave function at the energy of transformation ${\cal E}^{(2)}$
\label{sec.3.4}}

Now we shall find how WF at the energy of transformation ${\cal E}^{(2)}$ is deformed. If to use the partial solution of $\varphi_{w_{2}}^{(2)}$ at $\varphi_{w_{2}}^{(2)} = \phi_{w_{2}}^{(2)}$, then according to (\ref{eq.3.1.3}) and (\ref{eq.3.1.6}), the general solution for the deformed WF for potential $V_{1}^{\rm def}$ at such energy can be found so:
\begin{equation}
\begin{array}{cc}
  \varphi_{w_{2}}^{(1,\:{\rm def})} = \displaystyle\frac{I_{w_{2}}^{(2)} + \lambda_{2}} {\phi_{w_{2}}^{(2)}}, &
  I_{w_{2}}^{(2)} = \displaystyle\int\limits_{x_{0}}^{x} \Bigl(\phi_{w_{2}}^{(2)} (x')\Bigr)^{2} dx',
\end{array}
\label{eq.3.4.1}
\end{equation}
where, according to (\ref{eq.3.3.2}), we have:
\begin{equation}
  \phi_{w_{2}}^{(2)} =
    N_{w_{2}}^{(1)}
      \displaystyle\frac{
          \phi_{w_{1}}^{(1)} \phi_{w_{2}}^{(1),\prime} -
          \phi_{w_{1}}^{(1),\prime} \phi_{w_{2}}^{(1)}}
        {\phi_{w_{1}}^{(1)}}.
\label{eq.3.4.2}
\end{equation}
Calculating the integral:
\begin{equation}
\begin{array}{ccl}
  I_{w_{2}}^{(2)} (x) & = &
    \displaystyle\int\limits_{x_{0}}^{x}
    \Bigl( \phi_{w_{2}}^{(2)}(x') \Bigr)^{2} dx'=

  \Bigl( N_{w_{2}}^{(1)} \Bigr)^{2}
    \displaystyle\int\limits_{x_{0}}^{x}
      \displaystyle\frac{
          \Bigl(
            \phi_{w_{1}}^{(1)} \phi_{w_{2}}^{(1),\prime} -
            \phi_{w_{1}}^{(1),\prime} \phi_{w_{2}}^{(1)}
          \Bigr)^{2}}
        {\Bigl( \phi_{w_{1}}^{(1)} \Bigr)^{2}}  dx' = \\

  & = &
    \Bigl( N_{w_{2}}^{(1)} \Bigr)^{2}
      \biggl( \displaystyle\frac
        {\phi_{w_{2}}^{(1)}} {\phi_{w_{1}}^{(1)}}
      \biggr)
      \Bigl(
        \phi_{w_{1}}^{(1)} \phi_{w_{2}}^{(1),\prime} -
        \phi_{w_{1}}^{(1),\prime} \phi_{w_{2}}^{(1)}
      \Bigr) \Bigg|^{x'=x}_{x'=x_{0}} +

    \Bigl( N_{w_{2}}^{(1)} \Bigr)^{2}
      \bigl( {\cal E}^{(2)} - {\cal E}^{(1)} \bigr)
      \displaystyle\int\limits_{x_{0}}^{x}
        \Bigl(\phi_{w_{2}}^{(1)} \Bigr)^{2}  dx'
\end{array}
\label{eq.3.4.3}
\end{equation}
we obtain the deformed WF:
\begin{equation}
\begin{array}{ccl}
  \varphi_{w_{2}}^{(1,\:{\rm def})}(x) & = &
    N_{w_{2}}^{(1)} \phi_{w_{2}}^{(1)}(x) -

    N_{w_{2}}^{(1)}
    \displaystyle\frac
      {\phi_{w_{1}}^{(1)}(x) \phi_{w_{2}}^{(1)} (x_{0})}
      {\phi_{w_{1}}^{(1)} (x_{0})}
    \displaystyle\frac
      {\phi_{w_{1}}^{(1)}(x_{0}) \phi_{w_{2}}^{(1),\prime}(x_{0}) -
       \phi_{w_{1}}^{(1),\prime}(x_{0}) \phi_{w_{2}}^{(1)}(x_{0}) }
      {\phi_{w_{1}}^{(1)}(x) \phi_{w_{2}}^{(1),\prime}(x) -
       \phi_{w_{1}}^{(1),\prime}(x) \phi_{w_{2}}^{(1)}(x) } + \\

  & + &
    N_{w_{2}}^{(1)} \Bigl({\cal E}^{(2)} - {\cal E}^{(1)}\Bigr)
    \displaystyle\frac
      {\phi_{w_{1}}^{(1)}(x) }
      {\phi_{w_{1}}^{(1)}(x) \phi_{w_{2}}^{(1),\prime}(x) -
       \phi_{w_{1}}^{(1),\prime} \phi_{w_{2}}^{(1)}(x) }
      \displaystyle\int\limits_{x_{0}}^{x}
        \Bigl(\phi_{w_{2}}^{(1)}(x') \Bigr)^{2}  dx' +\\

  & + &
    \displaystyle\frac
      {\lambda_{2} \: \phi_{w_{1}}^{(1)}(x) }
      {N_{w_{2}}^{(1)}
       \Bigl(
         \phi_{w_{1}}^{(1)}(x) \phi_{w_{2}}^{(1),\prime}(x) -
         \phi_{w_{1}}^{(1),\prime}(x) \phi_{w_{2}}^{(1)} (x)
       \Bigr)}.
\end{array}
\label{eq.3.4.4}
\end{equation}

Now we take into account, that the solution of $\varphi_{w_{2}}^{(2)}$ can differ from the function of transformation $\phi_{w_{2}}^{(2)}$. In such a case, one can find non-zero function $\varphi_{w_{2}}^{(1,\:{\rm def})}$ so:
\begin{equation}
  \varphi_{w_{2}}^{(1,\:{\rm def})} =
    N_{w_{2}}^{(2)} A_{w_{2}}^{+} \varphi_{w_{2}}^{(2)} =
    N_{w_{2}}^{(2)}
      \biggl( -\displaystyle\frac{d}{dx} + W^{(2)} \biggr)
      \varphi_{w_{2}}^{(2)} =
    N_{w_{2}}^{(2)}
      \displaystyle\frac{
          -\phi_{w_{2}}^{(2)} \varphi_{w_{2}}^{(2),\prime} +
          \phi_{w_{2}}^{(2),\prime} \varphi_{w_{2}}^{(2)} }
        {\phi_{w_{2}}^{(2)}}.
\label{eq.3.4.5}
\end{equation}
Taking into account (\ref{eq.3.3.2}) and (\ref{eq.3.3.3}), we calculate the function $\varphi_{w_{2}}^{(2)}$ and its derivative:
\begin{equation}
\begin{array}{ccl}
  \varphi_{w_{2}}^{(2)} & = &
    N_{w_{2}}^{(1)}
      \displaystyle\frac{
          \phi_{w_{1}}^{(1)} \varphi_{w_{2}}^{(1),\prime} -
          \phi_{w_{1}}^{(1),\prime} \varphi_{w_{2}}^{(1)} }
        {\phi_{w_{1}}^{(1)}}, \\

  \varphi_{w_{2}}^{(2), \prime} & = &
   N_{w_{2}}^{(1)}
      \displaystyle\frac{
          \Bigl({\cal E}^{(1)} - {\cal E}^{(2)} \Bigr)
          \phi_{w_{1}}^{(1)}
          \phi_{w_{1}}^{(1)}
          \varphi_{w_{2}}^{(1)} -
          \phi_{w_{1}}^{(1)}
          \phi_{w_{1}}^{(1),\prime}
          \varphi_{w_{2}}^{(1),\prime} +
          \phi_{w_{1}}^{(1),\prime}
          \phi_{w_{1}}^{(1),\prime}
          \varphi_{w_{2}}^{(1)} }
        {\Bigl(\phi_{w_{1}}^{(1)}\Bigr)^{2}}
\end{array}
\label{eq.3.4.6}
\end{equation}
and obtain the deformed WF $\varphi_{w_{2}}^{(1,\:{\rm def})}$:
\begin{equation}
  \varphi_{w_{2}}^{(1,\:{\rm def})} =
    N_{w_{2}}^{(1)} N_{w_{2}}^{(2)}
    \Bigl({\cal E}^{(1)} - {\cal E}^{(2)} \Bigr)\:
    \displaystyle\frac
        {\phi_{w_{2}}^{(1)} \varphi_{w_{2}}^{(1),\prime} -
          \phi_{w_{2}}^{(1),\prime} \varphi_{w_{2}}^{(1)} }
        {\phi_{w_{1}}^{(1)} \phi_{w_{2}}^{(1),\prime} -
          \phi_{w_{1}}^{(1),\prime} \phi_{w_{2}}^{(1)} }\:
    \phi_{w_{1}}^{(1)}.
\label{eq.3.4.7}
\end{equation}
In the obtained solution one can separate the deformation explicitly:
\begin{equation}
  \varphi_{w_{2}}^{(1,\:{\rm def})} =
  N_{w_{2}}^{(1)} N_{w_{2}}^{(2)}
    \Bigl({\cal E}^{(2)} - {\cal E}^{(1)} \Bigr)\:
    \varphi_{w_{2}}^{(1)} +
  N_{w_{2}}^{(1)} N_{w_{2}}^{(2)}
    \Bigl({\cal E}^{(1)} - {\cal E}^{(2)} \Bigr)\:
    \displaystyle\frac
        {\phi_{w_{1}}^{(1)} \varphi_{w_{2}}^{(1),\prime} -
          \phi_{w_{1}}^{(1),\prime} \varphi_{w_{2}}^{(1)} }
        {\phi_{w_{1}}^{(1)} \phi_{w_{2}}^{(1),\prime} -
          \phi_{w_{1}}^{(1),\prime} \phi_{w_{2}}^{(1)} }\:
    \phi_{w_{2}}^{(1)}.
\label{eq.3.4.8}
\end{equation}

\section{Example --- deformation of rectangular well with finite width and infinitely high walls
\label{sec.example.rectangular}}

Let's consider, how the double complex SUSY-transformations are effective in construction of new real potentials on the basis of the given one. We shall analyze two cases: potential with discrete and potential wit continuous energy spectra.

As the starting potential $V_{1}$ with the discrete energy spectrum, we shall use the rectangular well with finite width and infinitely high walls:
\begin{equation}
\begin{array}{lcl}
  V_{1}(x) =
  \left \{
  \begin{array}{cl}
    0,       & \mbox{at } 0<x<a, \\
    +\infty, & \mbox{at } x<0 \mbox{ or } x>a.
  \end{array}
  \right.
\end{array}
\label{eq.example.1.1}
\end{equation}
If to restrict ourselves for this potential only the discrete values of energy higher then the wall bottom, then we have a general solution for WF for arbitrary state at arbitrary energy $E_{k}$:
\begin{equation}
\begin{array}{ll}
  \mbox{unbound state with index $k$:} &   \varphi_{k}^{(1)}(x) = A \sin(kx + \theta), \\
  \mbox{bound state with number $n$:} &  \varphi_{n}^{(1)}(x) =  \sqrt{\displaystyle\frac{2}{a}} \sin(k_{n} x)
\end{array}
\label{eq.example.1.2}
\end{equation}
and the discrete energy spectrum:
\begin{equation}
\begin{array}{ccc}
  E_{n} = E_{1} n^{2} = k_{n}^{2} =
    \displaystyle\frac{\pi^{2}}{a^{2}} n^{2}, &
  k_{n} = k_{1} n = \displaystyle\frac{\pi}{a} n, &
  n = 1, 2 \ldots
\end{array}
\label{eq.example.1.3}
\end{equation}
where 
$A$ and $\theta$ are arbitrary (complex) constants, WF of the bound state is normalized (is real, with accuracy to constant phase factor). Coordinates of \emph{nodes} for WF of the bound state with number $n$ equal to:
\begin{equation}
\begin{array}{cc}
  x_{l} = \displaystyle\frac{\pi l}{k_{n}} =  \displaystyle\frac{al}{n}, &
  l = 0, 1, 2 \ldots n
\end{array}
\label{eq.example.1.4}
\end{equation}
and are located at \emph{the same} distances, which are decreased with increasing of $n$.

Now if to include into analysis the complex values of the energies for the potential (\ref{eq.example.1.1}), then we obtain the general solution of the wave function concerning arbitrary selected complex energy ${\cal E}$:
\begin{equation}
\begin{array}{ccl}
  \varphi_{2}(x) = \alpha\,e^{cx} + \beta\,e^{-cx} = \alpha\,e^{\chi x} e^{ikx} + \beta\,e^{-\chi x} e^{-ikx}, & c = \chi + ik,
\end{array}
\label{eq.example.1.5}
\end{equation}
where $\alpha$ and $\beta$ are arbitrary complex constants, $c$ is \emph{``complex wave vector''}, $\chi=\Re{(c)}$ and $k=\Im{(c)}$ are real and imaginary parts of constant $c$. Substituting this solution into the Schr\"{o}dinger equation, we find the coefficients $\chi$ and $k$:
\begin{equation}
\begin{array}{ccc}
  \bigl(-c^{2} = {\cal E} \bigr) & \to &
  \begin{array}{cc}
    \Bigl\{ k^{2} - \chi^{2} = \Re{(\cal E)}, & -2\chi k = \Im{(\cal E)} \Bigl\}.
  \end{array}
\end{array}
\label{eq.example.1.6}
\end{equation}
If $\Im{({\cal E})} \ne 0$, then from (\ref{eq.example.1.6}) we obtain:
\begin{equation}
\begin{array}{cc}
  \Re{(c)}^{2} = \displaystyle\frac{1}{2} \:
        \Bigl( -\Re{({\cal E})} + \sqrt{ \Re{({\cal E})}^{2} + \Im{({\cal E})}^{2} } \Bigr), &
  \Im{(c)}^{2} = \displaystyle\frac{1}{2} \:
        \Bigl( \Re{({\cal E})} + \sqrt{ \Re{({\cal E})}^{2} + \Im{({\cal E})}^{2} } \Bigr).
\end{array}
\label{eq.example.1.7}
\end{equation}
If $\Im{({\cal E})}=0$, then:
\begin{equation}
\begin{array}{ll}
  \chi = \Re{(c)} = 0, & \mbox{\rm at  } {\cal E} = \Re{({\cal E})} > 0; \\
  k = \Im{(c)} = 0, & \mbox{\rm at  } {\cal E} = \Re{({\cal E})} < 0.
\end{array}
\label{eq.example.1.8}
\end{equation}

Further, as the method of the deformation we shall use the SUSY-approach with not coincident energies of transformation ${\cal E}^{(1)} \ne {\cal E}^{(2)}$. As two functions of transformation, we select the following functions:
\begin{equation}
\begin{array}{ccl}
  \phi_{1}(x) & = & e^{c_{1} x + \delta}; \\
  \phi_{2}(x) & = & \alpha\,e^{c_{2} x} + \beta\,e^{-c_{2}x},
\end{array}
\label{eq.example.1.9}
\end{equation}
where $\delta$ is arbitrary complex constant and $c_{1}$, $c_{2}$ are complex constants, defined by the complex functions of transformation ${\cal E}^{(1)}$, ${\cal E}^{(2)}$.

\subsection{Deformation of the potential
\label{sec.example.2}}

Lets' find, how the potential $V_{1}$ is deformed. According to (\ref{eq.3.2.1}), the potential after deformation has a form:
\begin{equation}
\begin{array}{ccl}
  V_{1}^{\rm (def)} (c_{w_{1}},c_{w_{2}}) & = &
    V_{1} + \Delta V_{1}(c_{w_{1}},c_{w_{2}}), \\
  \Delta V_{1} (c_{w_{1}},c_{w_{2}}) & = &
    2 \: \Bigl( {\cal E}^{(2)} - {\cal E}^{(1)} \Bigr) \,
      \displaystyle\frac{d}{dx} \displaystyle\frac
      {\phi_{w_{1}}^{(1)} \phi_{w_{2}}^{(1)}}
      {\phi_{w_{1}}^{(1)} \Bigl(\phi_{w_{2}}^{(1)}\Bigr)^{\prime} -
        \Bigl(\phi_{w_{1}}^{(1)} \Bigr)^{\prime} \phi_{w_{2}}^{(1)}}.
\end{array}
\label{eq.example.2.1}
\end{equation}
Substituting here solutions (\ref{eq.example.1.9}) for the functions of transformation, we obtain:
\begin{equation}
\begin{array}{ccl}
  \Delta V_{1} (c_{w_{1}},c_{w_{2}}) & = &
    2 \Bigl( {\cal E}^{(2)} - {\cal E}^{(1)} \Bigr)
      \displaystyle\frac{d}{dx} \displaystyle\frac
      {\alpha\,e^{c_{2} x} + \beta\,e^{-c_{2} x}}
      {\alpha\,(c_{2}-c_{1})\,e^{c_{2} x} - \beta\,(c_{2} + c_{1})\,e^{-c_{2} x}}.
\end{array}
\label{eq.example.2.2}
\end{equation}

The new potential $V_{1}^{\rm (def)}$ is complex. We shall be interesting in a question, whether the complex double SUSY-transformations proposed in this paper can give new real deformations of the potential $V_{1}$. The deformation $\Delta V_{1}$ to be real, it is enough, that the function is real (we shall find only nonconstant (and nonzero) deformations $\Delta V$):
\begin{equation}
\begin{array}{ccl}
  f (x) & = &
    \Bigl( {\cal E}^{(2)} - {\cal E}^{(1)} \Bigr)
    \bigl(\alpha\,e^{c_{2} x} + \beta\,e^{-c_{2} x} \bigr)\,
    \Bigl(\alpha^{*}\,(c_{2}^{*}-c_{1}^{*})\,e^{c_{2}^{*} x} -
          \beta^{*}\,(c_{2}^{*} + c_{1}^{*})\,e^{-c_{2}^{*} x}\Bigr).
\end{array}
\label{eq.example.2.3}
\end{equation}
Rewrite the function $f(x)$ explicitly:
\begin{equation}
\begin{array}{ccl}
  f (x) & = &
    \Bigl( {\cal E}^{(2)} - {\cal E}^{(1)} \Bigr)
    \Bigl( |\alpha|^{2}(c_{2}^{*}-c_{1}^{*})\,e^{(c_{2}+c_{2}^{*}) x} -
           |\beta|^{2}(c_{2}^{*}+c_{1}^{*})\,e^{-(c_{2}+c_{2}^{*}) x} + \\
& + &
           \alpha^{*}\,\beta\,(c_{2}^{*}-c_{1}^{*})\,e^{(c_{2}^{*}-c_{2}) x} -
           \alpha\,\beta^{*}\,(c_{2}^{*}+c_{1}^{*})\,e^{(c_{2}-c_{2}^{*}) x} \Bigr).
\end{array}
\label{eq.example.2.4}
\end{equation}
Taking into account:
\begin{equation}
\begin{array}{cc}
  c_{2} + c_{2}^{*} = 2\, \Re{(c_{2})}, &
  c_{2} - c_{2}^{*} = 2 i\, \Im{(c_{2})},
\end{array}
\label{eq.example.2.5}
\end{equation}
we obtain:
\begin{equation}
\begin{array}{ccl}
  f (x) & = &
    \Bigl( {\cal E}^{(2)} - {\cal E}^{(1)} \Bigr) \cdot
    \Bigl( |\alpha|^{2}(c_{2}^{*}-c_{1}^{*})\,e^{2\,\Re{(c_{2})} x} -
           |\beta|^{2}(c_{2}^{*}+c_{1}^{*})\,e^{-2\,\Re{(c_{2})} x} + \\
  & + &
           \alpha^{*}\,\beta\,(c_{2}^{*}-c_{1}^{*})\,e^{-2 i\,\Im{(c_{2})} x} -
           \alpha\,\beta^{*}\,(c_{2}^{*}+c_{1}^{*})\,e^{2 i\,\Im{(c_{2})} x} \Bigr).
\end{array}
\label{eq.example.2.6}
\end{equation}
On the basis of exponents $\exp{\pm 2\Re{(c_{2})x}}$ and $\exp{\pm 2i\Im{(c_{2})x}}$ one can combine functions $\sin{(2\Im{(c_{2})x})}$, $\cos{(2\Im{(c_{2})x})}$, $\cosh{(2\Re{(c_{2})x})}$ and $\sinh{(2\Re{(c_{2}))x}}$. Write:
\begin{equation}
\begin{array}{l}
  \vspace{2mm}
  \bigl( {\cal E}^{(2)} - {\cal E}^{(1)} \bigr) \cdot
  \Bigl(
    |\alpha|^{2}(c_{2}^{*}-c_{1}^{*})\,e^{2\,\Re{(c_{2})} x} -
    |\beta|^{2}(c_{2}^{*}+c_{1}^{*})\,e^{-2\,\Re{(c_{2})} x}
  \Bigr) = \\
  = N_{1}\,\cosh{\bigl(2\,\Re{(c_{2})}x\bigr)} + N_{2}\,\sinh{\bigl(2\,\Re{(c_{2})}x\bigr)} =
  e^{2\,\Re{(c_{2})} x}  \displaystyle\frac{N_{1} + N_{2}}{2} +
  e^{-2\,\Re{(c_{2})} x} \displaystyle\frac{N_{1} - N_{2}}{2}, \\

  \bigl( {\cal E}^{(2)} - {\cal E}^{(1)} \bigr) \cdot
  \Bigl(
    \alpha^{*}\,\beta\,(c_{2}^{*}-c_{1}^{*})\,e^{-2 i\,\Im{(c_{2})} x} -
    \alpha\,\beta^{*}\,(c_{2}^{*}+c_{1}^{*})\,e^{2 i\,\Im{(c_{2})} x}
  \Bigr) = \\
  = M_{1} \,\cos{\bigl(2\,\Im{(c_{2})}x\bigr)} + M_{2} \,\sin{\bigl(2\,\Im{(c_{2})}x\bigr)} =
  e^{2i\,\Im{(c_{2})} x}  \biggl(\displaystyle\frac{M_{1}}{2} + \displaystyle\frac{M_{2}}{2i} \biggr) +
  e^{-2i\,\Im{(c_{2})} x} \biggl(\displaystyle\frac{M_{1}}{2} - \displaystyle\frac{M_{2}}{2i} \biggr),
\end{array}
\label{eq.example.2.7}
\end{equation}
where $N_{1}$, $N_{2}$, $M_{1}$ and $M_{2}$ are new real constants. At the same exponents the coefficients must be equal:
\begin{equation}
\begin{array}{ll}
  \bigl( {\cal E}^{(2)} - {\cal E}^{(1)} \bigr) \cdot
    |\alpha|^{2}(c_{2}^{*}-c_{1}^{*}) = \displaystyle\frac{N_{1} + N_{2}}{2}, &
  \bigl( {\cal E}^{(2)} - {\cal E}^{(1)} \bigr) \cdot
    \alpha^{*}\,\beta\,(c_{2}^{*}-c_{1}^{*}) = \displaystyle\frac{M_{1}}{2} - \displaystyle\frac{M_{2}}{2i}, \\
  -\bigl( {\cal E}^{(2)} - {\cal E}^{(1)} \bigr) \cdot
    |\beta|^{2}(c_{2}^{*}+c_{1}^{*}) = \displaystyle\frac{N_{1} - N_{2}}{2}, &
  -\bigl( {\cal E}^{(2)} - {\cal E}^{(1)} \bigr) \cdot
    \alpha\,\beta^{*}\,(c_{2}^{*}+c_{1}^{*}) = \displaystyle\frac{M_{1}}{2} + \displaystyle\frac{M_{2}}{2i}.
\end{array}
\label{eq.example.2.8}
\end{equation}
From here we find $N_{1}$, $N_{2}$, $M_{1}$ and $M_{2}$, expressing them through $\alpha$, $\beta$, $c_{1}$ and $c_{2}$:
\begin{equation}
\begin{array}{lcl}
  N_{1} & = &
    \bigl( {\cal E}^{(2)} - {\cal E}^{(1)} \bigr) \:
    \bigl( |\alpha|^{2}(c_{2}^{*}-c_{1}^{*}) - |\beta|^{2}(c_{2}^{*}+c_{1}^{*}) \bigr), \\
  N_{2} & = &
    \bigl( {\cal E}^{(2)} - {\cal E}^{(1)} \bigr) \:
    \bigl( |\alpha|^{2}(c_{2}^{*}-c_{1}^{*}) + |\beta|^{2}(c_{2}^{*}+c_{1}^{*}) \bigr), \\
  M_{1} & = &
    \bigl( {\cal E}^{(2)} - {\cal E}^{(1)} \bigr) \:
    \bigl( \alpha^{*}\,\beta\,(c_{2}^{*}-c_{1}^{*}) - \alpha\,\beta^{*}\,(c_{2}^{*}+c_{1}^{*}) \bigr), \\
  M_{2} & = & i\,
    \bigl( {\cal E}^{(2)} - {\cal E}^{(1)} \bigr) \:
    \bigl(-\alpha^{*}\,\beta\,(c_{2}^{*}-c_{1}^{*}) - \alpha\,\beta^{*}\,(c_{2}^{*}+c_{1}^{*})\bigr).
\end{array}
\label{eq.example.2.9}
\end{equation}
Taking into account only real values for $N_{1}$, $N_{2}$, $M_{1}$, $M_{2}$, we obtain condition of the real deformation of the potential:
\begin{equation}
\begin{array}{l}
  \Im{\Bigl(
    \bigl( {\cal E}^{(2)} - {\cal E}^{(1)} \bigr) \:
    \bigl( |\alpha|^{2}(c_{2}^{*}-c_{1}^{*}) - |\beta|^{2}(c_{2}^{*}+c_{1}^{*}) \bigr)\Bigr)} = 0, \\
  \Im{\Bigl(
    \bigl( {\cal E}^{(2)} - {\cal E}^{(1)} \bigr) \:
    \bigl( |\alpha|^{2}(c_{2}^{*}-c_{1}^{*}) + |\beta|^{2}(c_{2}^{*}+c_{1}^{*}) \bigr)\Bigr)} = 0, \\
  \Im{\Bigl(
    \bigl( {\cal E}^{(2)} - {\cal E}^{(1)} \bigr) \:
    \bigl( \alpha^{*}\,\beta\,(c_{2}^{*}-c_{1}^{*}) - \alpha\,\beta^{*}\,(c_{2}^{*}+c_{1}^{*}) \bigr)\Bigr)} = 0, \\
  \Re{\Bigl(
    \bigl( {\cal E}^{(2)} - {\cal E}^{(1)} \bigr) \:
    \bigl( \alpha^{*}\,\beta\,(c_{2}^{*}-c_{1}^{*}) + \alpha\,\beta^{*}\,(c_{2}^{*}+c_{1}^{*}) \bigr)\Bigr)} = 0.
\end{array}
\label{eq.example.2.10}
\end{equation}

Write:
\begin{equation}
\begin{array}{lcl}
  {\cal E}^{(2)} - {\cal E}^{(1)} =
  -\bigl(c_{2}\bigr)^{2} + \bigl(c_{1}\bigr)^{2} =
  -\Re{(c_{2})}^{2} + \Im{(c_{2})}^{2} + \Re{(c_{1})}^{2} - \Im{(c_{1})}^{2} +
  2i \cdot \Bigl(-\Re{(c_{2})}\Im{(c_{2})} + \Re{(c_{1})}\Im{(c_{1})} \Bigr).
\end{array}
\label{eq.example.2.11}
\end{equation}
Taking into account of this, we rewrite the first two equations of (\ref{eq.example.2.10}), separating real and imaginary parts:
\[
\begin{array}{l}
  \Re{\bigl( {\cal E}^{(2)} - {\cal E}^{(1)} \bigr)} \cdot
  \Bigl\{\bigl(\Re{(\alpha)}^{2}+\Im{(\alpha)}^{2} \bigr) \,\bigl(-\Im{(c_{2})}+\Im{(c_{1})}\bigr) -
         \bigl(\Re{(\beta)}^{2}+\Im{(\beta)}^{2} \bigr) \,\bigl(-\Im{(c_{2})}-\Im{(c_{1})}\bigr)
  \Bigr\} + \\
 + \Im{\bigl( {\cal E}^{(2)} - {\cal E}^{(1)} \bigr)} \cdot
  \Bigl\{\bigl(\Re{(\alpha)}^{2}+\Im{(\alpha)}^{2} \bigr) \,\bigl(\Re{(c_{2})}-\Re{(c_{1})}\bigr) -
         \bigl(\Re{(\beta)}^{2}+\Im{(\beta)}^{2} \bigr) \,\bigl(\Re{(c_{2})}+\Re{(c_{1})}\bigr)
  \Bigr\} = 0, \\

  \Re{\bigl( {\cal E}^{(2)} - {\cal E}^{(1)} \bigr)} \cdot
  \Bigl\{\bigl(\Re{(\alpha)}^{2}+\Im{(\alpha)}^{2} \bigr) \,\bigl(-\Im{(c_{2})}+\Im{(c_{1})}\bigr) +
         \bigl(\Re{(\beta)}^{2}+\Im{(\beta)}^{2} \bigr) \,\bigl(-\Im{(c_{2})}-\Im{(c_{1})}\bigr)
  \Bigr\} + \\
 + \Im{\bigl( {\cal E}^{(2)} - {\cal E}^{(1)} \bigr)} \cdot
  \Bigl\{\bigl(\Re{(\alpha)}^{2}+\Im{(\alpha)}^{2} \bigr) \,\bigl(\Re{(c_{2})}-\Re{(c_{1})}\bigr) +
         \bigl(\Re{(\beta)}^{2}+\Im{(\beta)}^{2} \bigr) \, \bigl(\Re{(c_{2})}+\Re{(c_{1})}\bigr)
  \Bigr\} = 0.
\end{array}
\]
Add and subtract one equation concerning the other one:
\begin{equation}
\begin{array}{l}
  \bigl(\Re{(\alpha)}^{2}+\Im{(\alpha)}^{2} \bigr) \cdot
  \Bigl\{
    \Re{\bigl( {\cal E}^{(2)} - {\cal E}^{(1)} \bigr)} \, \bigl(-\Im{(c_{2})}+\Im{(c_{1})}\bigr) +
    \Im{\bigl( {\cal E}^{(2)} - {\cal E}^{(1)} \bigr)} \, \bigl(\Re{(c_{2})}-\Re{(c_{1})} \bigr)
  \Bigr\} = 0, \\
  \bigl(\Re{(\beta)}^{2}+\Im{(\beta)}^{2} \bigr) \cdot \,
  \Bigr\{
    \Re{\bigl( {\cal E}^{(2)} - {\cal E}^{(1)} \bigr)} \, \bigl(-\Im{(c_{2})}-\Im{(c_{1})} \bigr) +
    \Im{\bigl( {\cal E}^{(2)} - {\cal E}^{(1)} \bigr)} \, \bigl( \Re{(c_{2})}+\Re{(c_{1})} \bigr)
   \Bigr\} = 0.
\end{array}
\label{eq.example.2.12}
\end{equation}
And obtain the first two conditions:
\begin{equation}
\begin{array}{ll}
\vspace{2mm}
  1) & \Re{\bigl( {\cal E}^{(2)} - {\cal E}^{(1)} \bigr)} \, \bigl(-\Im{(c_{2})}+\Im{(c_{1})}\bigr) +
       \Im{\bigl( {\cal E}^{(2)} - {\cal E}^{(1)} \bigr)} \, \bigl(\Re{(c_{2})}-\Re{(c_{1})} \bigr) = 0, \\
  2) & \Re{\bigl( {\cal E}^{(2)} - {\cal E}^{(1)} \bigr)} \:
       \bigl(-\Im{(c_{2})}-\Im{(c_{1})} \bigr) \, + \Im{\bigl( {\cal E}^{(2)} - {\cal E}^{(1)} \bigr)}
       \bigl(\Re{(c_{2})}+\Re{(c_{1})}\bigr) = 0.
\end{array}
\label{eq.example.2.13}
\end{equation}

If $\Im{({\cal E}_{2} - {\cal E}_{1})} = 0$, then from (\ref{eq.example.2.13}) we obtain:
\begin{equation}
\begin{array}{cccc}
  \Im{(c_{1})} = 0, & \Im{(c_{2})} = 0, & \hspace{5mm}  \Im{({\cal E}_{1})} = 0, & \Im{({\cal E}_{2})} = 0.
\end{array}
\label{eq.example.2.14}
\end{equation}

At $\Im{({\cal E}_{1})}=0$ and $\Im{({\cal E}_{2})}=0$ the exponents $\exp{\pm 2i\Im{(c_{2})} x}$ in (\ref{eq.example.2.6}) and (\ref{eq.example.2.7}) equal to 1. Therefore, instead of the third and the forth conditions of (\ref{eq.example.2.10}) we must use the following one:
\begin{equation}
  \Im{\bigl( \alpha^{*}\,\beta (c_{2}^{*}-c_{1}^{*}) - \alpha\,\beta^{*} (c_{2}^{*}+c_{1}^{*}) \bigr)} = 0.
\label{eq.example.2.15}
\end{equation}
We obtain:
\begin{equation}
\begin{array}{l}
  \Re{(\alpha)} \, \Im{(\beta)} = \Im{(\alpha)} \, \Re{(\beta)}.
\end{array}
\label{eq.example.2.16}
\end{equation}
{\bf One can see, that the coefficients $\alpha$ and $\beta$ must be real only (with accuracy to the same complex factor), that the deformation of the potential $\Delta V$ to be real.}

\vspace{7mm}
If $\Im{({\cal E}_{2} - {\cal E}_{1})} \ne 0$, then adding and subtracting one expression concerning the other one in (\ref{eq.example.2.13}), we obtain:
\begin{equation}
\begin{array}{ll}
  \Re{\bigl( {\cal E}^{(2)} - {\cal E}^{(1)} \bigr)} \, \Im{(c_{2})} -
    \Im{\bigl( {\cal E}^{(2)} - {\cal E}^{(1)} \bigr)} \, \Re{(c_{2})} = 0, \\
  \Re{\bigl( {\cal E}^{(2)} - {\cal E}^{(1)} \bigr)} \, \Im{(c_{1})} \bigr) -
    \Im{\bigl( {\cal E}^{(2)} - {\cal E}^{(1)} \bigr)} \, \Re{(c_{1})} = 0.
\end{array}
\label{eq.example.2.17}
\end{equation}
Substituting the real and imaginary parts of ${\cal E}_{2} - {\cal E}_{1}$ from (\ref{eq.example.2.11}), we find:
\begin{equation}
\begin{array}{ll}
\vspace{2mm}
  \Bigl[ -\Re{(c_{2})}^{2} + \Im{(c_{2})}^{2} + \Re{(c_{1})}^{2} - \Im{(c_{1})}^{2} \Bigr] \cdot \Im{(c_{2})} -
    2 \cdot \Bigl(-\Re{(c_{2})}\Im{(c_{2})} + \Re{(c_{1})}\Im{(c_{1})} \Bigr) \cdot \Re{(c_{2})} = \\
\vspace{2mm}
  = \bigl[ \Re{(c_{2})}^{2} + \Im{(c_{2})}^{2} + \Re{(c_{1})}^{2} - \Im{(c_{1})}^{2} \Bigr] \cdot \Im{(c_{2})} -
    2 \: \Re{(c_{1})}\, \Re{(c_{2})}\, \Im{(c_{1})} = 0, \\

\vspace{2mm}
  \Bigl[ -\Re{(c_{2})}^{2} + \Im{(c_{2})}^{2} + \Re{(c_{1})}^{2} - \Im{(c_{1})}^{2} \Bigr] \cdot \Im{(c_{1})} -
    2 \cdot \Bigl(-\Re{(c_{2})}\Im{(c_{2})} + \Re{(c_{1})}\Im{(c_{1})} \Bigr) \cdot \Re{(c_{1})} = \\
  = \Bigl[ -\Re{(c_{2})}^{2} + \Im{(c_{2})}^{2} - \Re{(c_{1})}^{2} - \Im{(c_{1})}^{2} \Bigr] \cdot \Im{(c_{1})} +
    2 \:\Re{(c_{1})}\,\Re{(c_{2})}\,\Im{(c_{2})} = 0.
\end{array}
\label{eq.example.2.18}
\end{equation}
Multiply the first equation on $\Im{(c_{2})}$, and the second one --- on $\Im{(c_{1})}$:
\begin{equation}
\begin{array}{ll}
\vspace{2mm}
  \Bigl[ \Re{(c_{2})}^{2} + \Im{(c_{2})}^{2} + \Re{(c_{1})}^{2} - \Im{(c_{1})}^{2} \Bigr] \, \Im{(c_{2})}^{2} -
    2 \:\Re{(c_{1})}\, \Re{(c_{2})}\, \Im{(c_{1})}\,\Im{(c_{2})} = 0, \\
  \Bigl[ -\Re{(c_{2})}^{2} + \Im{(c_{2})}^{2} - \Re{(c_{1})}^{2} - \Im{(c_{1})}^{2} \Bigr] \, \Im{(c_{1})}^{2} +
    2 \:\Re{(c_{1})}\,\Re{(c_{2})}\,\Im{(c_{1})}\,\Im{(c_{2})} = 0.
\end{array}
\label{eq.example.2.19}
\end{equation}
Adding one equation with the second, we obtain:
\[
\begin{array}{ll}
\vspace{2mm}
  \bigl[ \Re{(c_{1})}^{2} + \Re{(c_{2})}^{2} \bigr] \, \bigl( \Im{(c_{2})}^{2} - \Im{(c_{1})}^{2} \bigr) +
  \bigl[ \Im{(c_{2})}^{2} - \Im{(c_{1})}^{2} \bigr] \, \bigl( \Im{(c_{2})}^{2} + \Im{(c_{1})}^{2} \bigr) = \\
  = \;\Bigl[ \Re{(c_{1})}^{2} + \Re{(c_{2})}^{2} + \Im{(c_{2})}^{2} + \Im{(c_{1})}^{2} \Bigr] \cdot
  \bigl( \Im{(c_{2})}^{2} - \Im{(c_{1})}^{2} \bigr) = 0,
\end{array}
\]
from here we find:
\begin{equation}
  \Im{(c_{1})} = \pm\Im{(c_{2})}.
\label{eq.example.2.20}
\end{equation}
Now subtract the second equation from the first one in (\ref{eq.example.2.19}):
\begin{equation}
\begin{array}{ll}
  \bigl( \Re{(c_{1})}^{2} + \Re{(c_{2})}^{2} \bigr) \, \bigl( \Im{(c_{2})}^{2} + \Im{(c_{1})}^{2} \bigr) -
  4 \:\Re{(c_{1})}\,\Re{(c_{1})}\,\Im{(c_{1})}\,\Im{(c_{2})} = 0.
\end{array}
\label{eq.example.2.21}
\end{equation}
Taking into account (\ref{eq.example.2.20}), we obtain:
\begin{equation}
\begin{array}{lllll}
\vspace{2mm}
  1) & \Im{(c_{1})} = +\Im{(c_{2})}: &
    \Bigl( \Re{(c_{1})}^{2} + \Re{(c_{2})}^{2} - 2 \:\Re{(c_{1})}\,\Re{(c_{1})} =
    \bigl( \Re{(c_{1})} - \Re{(c_{2})} \bigr)^{2} = 0 \Bigr) & \to &
    \bigl( \Re{(c_{1})} = +\Re{(c_{2})} \bigr),  \\
  2) & \Im{(c_{1})} = -\Im{(c_{2})}: &
    \Bigl( \Re{(c_{1})}^{2} + \Re{(c_{2})}^{2} + 2 \:\Re{(c_{1})}\,\Re{(c_{1})} =
    \bigl( \Re{(c_{1})} + \Re{(c_{2})} \bigr)^{2} = 0 \Bigr) & \to &
    \bigl( \Re{(c_{1})} = -\Re{(c_{2})} \bigr).
\end{array}
\label{eq.example.2.22}
\end{equation}
And we come to absolute coincidence between two energies of transformation. Therefore, we cannot use such solutions and the case $\Im{({\cal E}_{2} - {\cal E}_{1})} \ne 0$.

\subsection{Deformation of wave function at arbitrary energy $E_{k}$, not coincident with the energies of transformation: $E_{k} \ne {\cal E}^{(1)}$ and $E_{k} \ne {\cal E}^{(2)}$
\label{sec.example.3}}

Let's the general solution for WF $\varphi_{k}^{(1)}$ at the energy $E_{k}$ (higher then the bottom of the well) before the deformation has a form:
\begin{equation}
  \varphi_{k}^{(1)} (x) = \alpha_{k}\,\sin{kx} + \beta_{k}\,\cos{kx},
\label{eq.example.3.1}
\end{equation}
where $\alpha_{k}$ and $\beta_{k}$ are arbitrary complex constants. According to (\ref{eq.3.1.11}), we find the form of this WF after the deformation:
\begin{equation}
\begin{array}{lcl}
  \varphi_{k}^{(1,\:{\rm def})} (x) & = &
  N_{k}^{(1)} N_{k}^{(2)} \cdot \Bigg\{ \Bigl( E_{k}^{(1)} - {\cal E}^{(1)} \Bigr)
    \bigl( \alpha_{k}\,\sin{kx} + \beta_{k}\,\cos{kx} \bigr) + \\
  & + & \Bigl( {\cal E}^{(1)} - {\cal E}^{(2)} \Bigr) \bigl( \alpha\,e^{c_{2}x} + \beta\,e^{-c_{2}x} \bigr) \,
    \displaystyle\frac
      {(k \alpha_{k} - c_{1}\beta_{k})\,\cos{kx} - (k \beta_{k} + c_{1}\alpha_{k})\,\sin{kx}}
      {(c_{2}-c_{1})\,\alpha\,e^{c_{2}x} - (c_{2}+c_{1})\,\beta\,e^{-c_{2}x}}
    \Bigg\}.
\end{array}
\label{eq.example.3.3}
\end{equation}

\subsection{Bound states and the energy spectrum
\label{sec.example.4}}

Now we consider the bound state with WF:
\begin{equation}
\begin{array}{cc}
  \varphi_{k}^{(1,\:{\rm def})} (x=0) = 0, &
  \varphi_{k}^{(1,\:{\rm def})} (x=a) = 0.
\end{array}
\label{eq.example.4.1}
\end{equation}
At point $x=0$ we have:
\begin{equation}
\begin{array}{lcl}
  \varphi_{k}^{(1,\:{\rm def})} (x=0) & = &
  N_{k}^{(1)} N_{k}^{(2)} \cdot
    \Bigg\{ \Bigl( E_{k}^{(1)} - {\cal E}^{(1)} \Bigr)\, \beta_{k} +
      \Bigl( {\cal E}^{(1)} - {\cal E}^{(2)} \Bigr) \bigl( \alpha + \beta \bigr)
      \displaystyle\frac{k\,\alpha_{k} - c_{1}\,\beta_{k}} {c_{2}\,\bigl( \alpha - \beta \bigr) -
      c_{1}\,\bigl( \alpha + \beta \bigr)}
    \Bigg\} = 0.
\end{array}
\label{eq.example.4.2}
\end{equation}
From here we find:
\begin{equation}
\begin{array}{lcl}
  \alpha_{k} \cdot k \:
    \displaystyle\frac{\bigl( {\cal E}^{(1)} - {\cal E}^{(2)} \bigr)\,(\alpha + \beta)}
    {c_{2}\,(\alpha - \beta) - c_{1}\,(\alpha + \beta)} =

  \beta_{k} \cdot \biggl[ {\cal E}^{(1)} - E_{k}^{(1)} +
     \displaystyle\frac{\bigl({\cal E}^{(1)} - {\cal E}^{(2)}\bigr)\,c_{1}\,(\alpha + \beta)}
     {c_{2}\,(\alpha - \beta) - c_{1}\,(\alpha + \beta)}
    \biggr].
\end{array}
\label{eq.example.4.3}
\end{equation}
At $\alpha+\beta=0$ we obtain:
\begin{equation}
\begin{array}{lcl}
  \beta_{k} = 0, & \varphi_{k}^{(1)} (x) = \alpha_{k}\,\sin{kx}.
\end{array}
\label{eq.example.4.4}
\end{equation}
At $\alpha+\beta \ne 0$ we obtain:
\begin{equation}
  \alpha_{k} = \beta_{k} \cdot T_{k},
\label{eq.example.4.5}
\end{equation}
where
\begin{equation}
\begin{array}{lcl}
  T_{k} & = &
  \displaystyle\frac{c_{1}}{k} \; + \;
  \displaystyle\frac{{\cal E}^{(1)} - E_{k}^{(1)}}{k}\,
  \displaystyle\frac{c_{2}\,(\alpha - \beta) - c_{1}\,(\alpha + \beta)}
                    {\bigl({\cal E}^{(1)} - {\cal E}^{(2)} \bigr)\,(\alpha + \beta)}.
\end{array}
\label{eq.example.4.6}
\end{equation}
Taking into account $E_{k} = k^{2}$, we rewrite (\ref{eq.example.4.6}) so:
\begin{equation}
\begin{array}{lcl}
  T_{k} & = &
  \displaystyle\frac{c_{1} + {\cal E}^{(1)}}{k} \cdot
  \displaystyle\frac{c_{2}\,(\alpha - \beta) - c_{1}\,(\alpha + \beta)}
                    {\bigl({\cal E}^{(1)} - {\cal E}^{(2)} \bigr)\,(\alpha + \beta)} -
  k \cdot
  \displaystyle\frac{c_{2}\,(\alpha - \beta) - c_{1}\,(\alpha + \beta)}
                            {\bigl({\cal E}^{(1)} - {\cal E}^{(2)} \bigr)\,(\alpha + \beta)}.
\end{array}
\label{eq.example.4.7}
\end{equation}
According to the previous section, if we consider only the real deformations of the rectangular well, thet the function:
\begin{equation}
\begin{array}{lcl}
  S(x) & = &
    \Bigl({\cal E}^{(1)} - {\cal E}^{(2)}\Bigr) \,
    \displaystyle\frac
      {\alpha\,e^{c_{2}x} + \beta\,e^{-c_{2}x}}
      {\alpha\,(c_{2}-c_{1})\,e^{c_{2}x} - \beta\,(c_{2}+c_{1})\,e^{-c_{2}x}}
\end{array}
\label{eq.example.4.8}
\end{equation}
becomes real (at $x \in [0, a]$). Taking into account, that it does not depend on $k$, we rewrite (\ref{eq.example.4.7}) so:
\begin{equation}
  T_{k} = \displaystyle\frac{c_{1} + {\cal E}^{(1)}}{S(0)\,k} - \displaystyle\frac{k} {S(0)}.
\label{eq.example.4.9}
\end{equation}

Now consider the deformed WF at the second point $x=a$:
\begin{equation}
\begin{array}{lcl}
  \varphi_{k}^{(1,\:{\rm def})} (x=a) & = &
  N_{k}^{(1)} N_{k}^{(2)} \cdot \Bigg\{ \Bigl( E_{k}^{(1)} - {\cal E}^{(1)} \Bigr)
    \bigl( \alpha_{k}\,\sin{ka} + \beta_{k}\,\cos{ka} \bigr) + \\
\vspace{1mm}
  & + & \Bigl( {\cal E}^{(1)} - {\cal E}^{(2)} \Bigr) \bigl( \alpha\,e^{c_{2}a} + \beta\,e^{-c_{2}a} \bigr) \,
    \displaystyle\frac
      {(k \alpha_{k} - c_{1}\beta_{k})\,\cos{ka} - (k \beta_{k} + c_{1}\alpha_{k})\,\sin{ka}}
      {(c_{2}-c_{1})\,\alpha\,e^{c_{2}a} - (c_{2}+c_{1})\,\beta\,e^{-c_{2}a}}
    \Bigg\} = 0.
\end{array}
\label{eq.example.4.10}
\end{equation}
From here we find:
\begin{equation}
\begin{array}{lcl}
  \sin{ka} \cdot
    \biggl\{ -\alpha_{k}\,\Bigl( E_{k}^{(1)} - {\cal E}^{(1)} \Bigr) + S(a) \: (k \beta_{k} + c_{1}\alpha_{k}) \biggr\} =
  \cos{ka} \cdot
    \biggl\{ \beta_{k}\,\Bigl( E_{k}^{(1)} - {\cal E}^{(1)} \Bigr) +  S(a) \:(k \alpha_{k} - c_{1}\beta_{k}) \biggr\}.
\end{array}
\label{eq.example.4.11}
\end{equation}
Taking into account (\ref{eq.example.4.5}), we obtain:
\begin{equation}
  \sin{ka} \cdot \biggl\{ T_{k}\,\Bigl( {\cal E}^{(1)} - k^{2} \Bigr) + S(a) \: (k + c_{1} T_{k}) \biggr\} =
  \cos{ka} \cdot \biggl\{ k^{2} - {\cal E}^{(1)} +  S(a) \:(k T_{k} - c_{1}) \biggr\}.
\label{eq.example.4.12}
\end{equation}
Substituting $T_{k}$ from (\ref{eq.example.4.9}), we find:
\begin{equation}
\begin{array}{lcl}
  \sin{ka} \cdot
    \Biggl\{
      \Bigl[ c_{1} + {\cal E}^{(1)} - k^{2} \Bigr]\,
      \Bigl( {\cal E}^{(1)} - k^{2} \Bigr) +
        S(0)\,S(a)\,k^{2} + S(a)\,c_{1}\,\Bigl[ c_{1} + {\cal E}^{(1)} - k^{2}
      \Bigr]
      \Biggr\} = \\
  = \cos{ka} \cdot k \cdot
    \biggl\{
      S(0)\,\bigl( k^{2} - {\cal E}^{(1)} \bigr) +
      S(a)\,\bigl( c_{1} + {\cal E}^{(1)} - k^{2} \bigr) -
      S(0)\,S(a)\,c_{1}
    \biggr\}.
\end{array}
\label{eq.example.4.13}
\end{equation}

Now we have obtained equation of the energy spectrum for the deformed potential. At $\cos{ka} \ne 0$ one can write:
\begin{equation}
  \tan{ka} = \displaystyle\frac{k\,(A + B\,k^{2})}{C + D\,k^{2} + k^{4}},
\label{eq.example.4.14}
\end{equation}
where
\begin{equation}
\begin{array}{lcl}
\vspace{2mm}
  A & = & -S(0)\,{\cal E}^{(1)} + S(a)\,\bigl( c_{1} + {\cal E}^{(1)} \bigr) - S(0)\,S(a)\,c_{1}, \\
\vspace{2mm}
  B & = & S(0) - S(a), \\
\vspace{2mm}
  C & = &
      \bigl( c_{1} + {\cal E}^{(1)} \bigr)\, {\cal E}^{(1)} + S(a)\,c_{1}\, \bigl( c_{1} + {\cal E}^{(1)} \bigr), \\
\vspace{2mm}
  D & = & -c_{1} - 2{\cal E}^{(1)} - S(a)\,c_{1} + S(0)\,S(a).
\end{array}
\label{eq.example.4.15}
\end{equation}
At
\begin{equation}
\begin{array}{cc}
  A = 0, & B = 0
\end{array}
\label{eq.example.4.16}
\end{equation}
we come to the energy spectrum for the rectangular well --- before its deformation. For this, the following condition must be fulfilled:
\begin{equation}
\begin{array}{cc}
  S(a) \cdot (1-S(0)) \cdot c_{1} = 0, &  \hspace{10mm} S(0) = S(a).
\end{array}
\label{eq.example.4.17}
\end{equation}

\section{Deformation of potentials with continuous energy spectra and new reflectionless potentials
\label{sec.example.reflectionless}}

Now we shall analyze whether the double complex SUSY-transformations can give some-thing interesting in construction of new real potentials with continuous energy spectra. Let's consider the simplest case --- deformation of null potential. Omitting external walls of the rectangular well (\ref{eq.example.1.1}), we transform it into such null potential. Here, one can use formulas, obtained in the previous section.

As before, we use the functions of transformation in the form (\ref{eq.example.1.9}). The deformed null potential $V_{1}^{\rm (def)}$ has the form:
\begin{equation}
\begin{array}{ccl}
  V_{2} (x) & = &
    2 \Bigl( {\cal E}^{(2)} - {\cal E}^{(1)} \Bigr)
      \displaystyle\frac{d}{dx} \displaystyle\frac
      {\alpha\,e^{c_{2} x} + \beta\,e^{-c_{2} x}}
      {\alpha\,(c_{2}-c_{1})\,e^{c_{2} x} - \beta\,(c_{2} + c_{1})\,e^{-c_{2} x}}.
\end{array}
\label{eq.example.reflectionless.1}
\end{equation}
This potential is complex. To be it real, we use real negative ${\cal E}^{(1)}$ and ${\cal E}^{(2)}$, and we use the conditions (\ref{eq.example.2.14}), (\ref{eq.example.2.16}):
\begin{equation}
\begin{array}{cccc}
  \Im{(c_{1})} = 0, & \Im{(c_{2})} = 0, & \hspace{5mm} &
  \Re{(\alpha)} \, \Im{(\beta)} = \Im{(\alpha)} \, \Re{(\beta)}.
\end{array}
\label{eq.example.reflectionless.2}
\end{equation}

Lets' analyze behavior of the potential $V_{2}$ in asymptotic regions. At limits $x \to \pm\infty$ we obtain (in numerator and denominator we write only maximal components):
\begin{equation}
  V_{2} (x\to \pm\infty) \; = \;
    \displaystyle\frac
      {g_{1}\, e^{2\Re{(c_{2})} x} + g_{2}\, e^{-2\Re{(c_{2})} x}}
      { |\alpha|^{4} |c_{2}-c_{1}|^{4} e^{4\Re{(c_{2})} x} + |\beta|^{4}  |c_{2}+c_{1}|^{4} e^{-4\Re{(c_{2})} x}}
    \Bigg|_{x \to \pm\infty} = 0,
\label{eq.example.reflectionless.3}
\end{equation}
where $g_{1}$ and $g_{2}$ are constants. So, in both asymptotic regions $V_{2}$ tends to zero. Therefore, consideration of propagation of wave in this potential is a natural problem, and general solution for wave function for such wave in both asymptotic regions is linear combination of two plane waves $\exp{(\pm ikx)}$, with opposite directions of propagation.

Now we clarify whether $V_{2}$ has divergencies and discontinuities. From (\ref{eq.example.reflectionless.1}) we see that this potential is continuous if denominator under differential in (\ref{eq.example.reflectionless.1}) is not equal to zero. So, the condition of existence of divergency at point $x_{d}$ is the following:
\begin{equation}
\begin{array}{ccl}
  \alpha\,(c_{2}-c_{1})\,e^{c_{2} x_{d}} - \beta\,(c_{2} + c_{1})\,e^{-c_{2} x_{d}} = 0.
\end{array}
\label{eq.example.reflectionless.4}
\end{equation}
We find coordinate of divergency:
\begin{equation}
  x_{d} =
    \displaystyle\frac{1}{2c_{2}}
    \log {\biggl(\displaystyle\frac{\beta\,(c_{2} + c_{1})} {\alpha\,(c_{2}-c_{1})} \biggr)}.
\label{eq.example.reflectionless.6}
\end{equation}
Using formula for logarithm with complex argument:
\begin{equation}
  \log{z} = \log |z| + i \arg{z},
\label{eq.example.reflectionless.7}
\end{equation}
rewrite coordinate (\ref{eq.example.reflectionless.6}) so:
\begin{equation}
  x_{d} =
    \displaystyle\frac{1}{2c_{2}}
    \log {\biggl|\displaystyle\frac{\beta\,(c_{2} + c_{1})} {\alpha\,(c_{2}-c_{1})} \biggr|} +
    \displaystyle\frac{i}{2c_{2}}
    \arg {\biggl(\displaystyle\frac{\beta\,(c_{2} + c_{1})} {\alpha\,(c_{2}-c_{1})} \biggr)}.
\label{eq.example.reflectionless.8}
\end{equation}
Taking into account that $x_{d}$ can be real only, we see that at arbitrary values of $c_{1}$, $c_{2}$, $\alpha$, $\beta$ the expression (\ref{eq.example.reflectionless.8}) is not correct. For example, if to use:
\begin{equation}
\begin{array}{cc}
  \displaystyle\frac{\beta\,(c_{2} + c_{1})} {\alpha\,(c_{2}-c_{1})} \in \Re, &
  \displaystyle\frac{\beta\,(c_{2} + c_{1})} {\alpha\,(c_{2}-c_{1})} < 0,
\end{array}
\label{eq.example.reflectionless.9}
\end{equation}
then the potential (\ref{eq.example.reflectionless.1}) has not divergencies and is continuous in the whole axis $x$. So, we have proved that with exception of some values of $c_{1}$, $c_{2}$, $\alpha$, $\beta$, the potential $V_{2}(x)$ is continuous function in the whole axis $x$.

\begin{figure}[htbp]
\centerline{
\includegraphics[width=74mm]{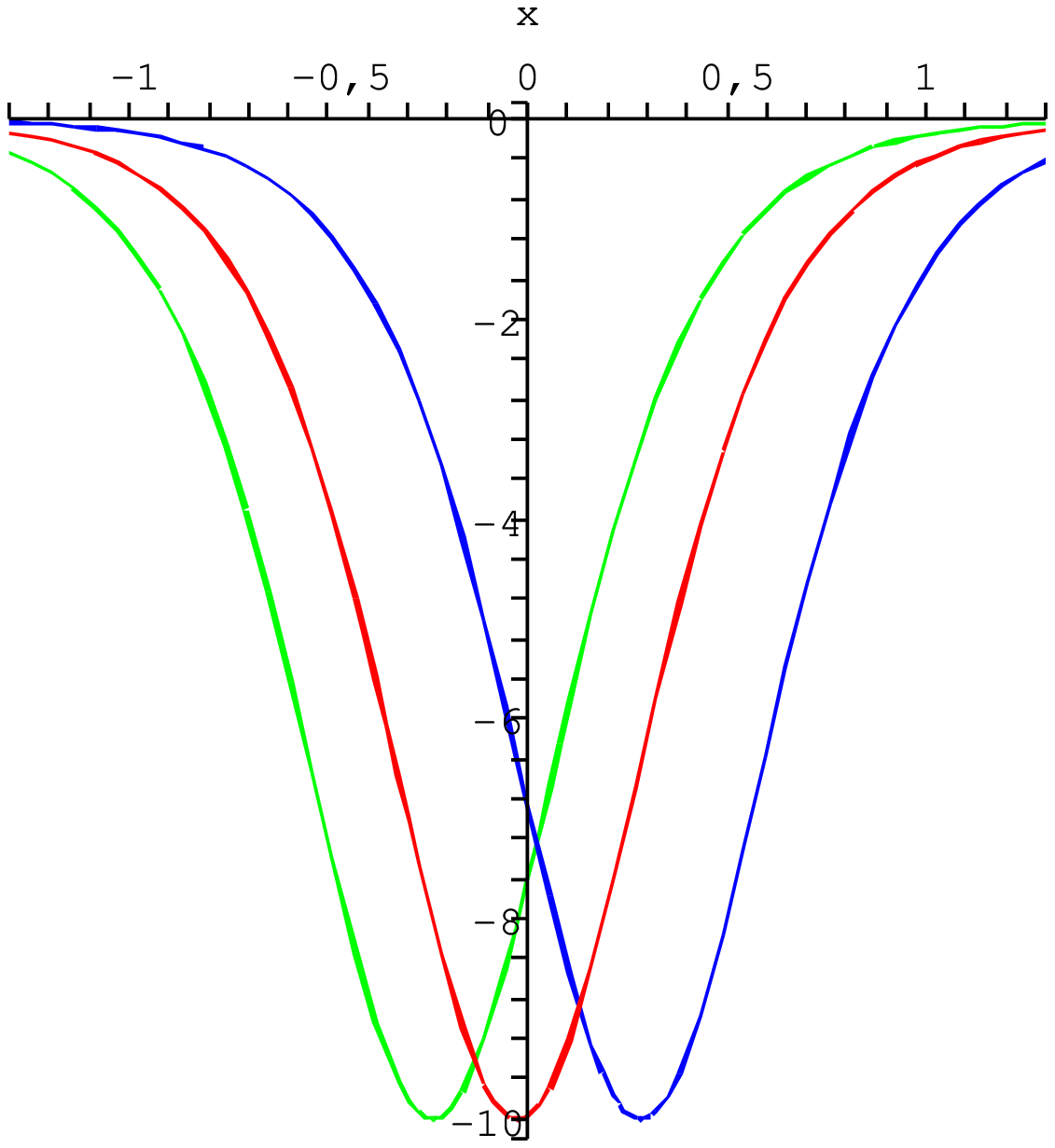}
\includegraphics[width=74mm]{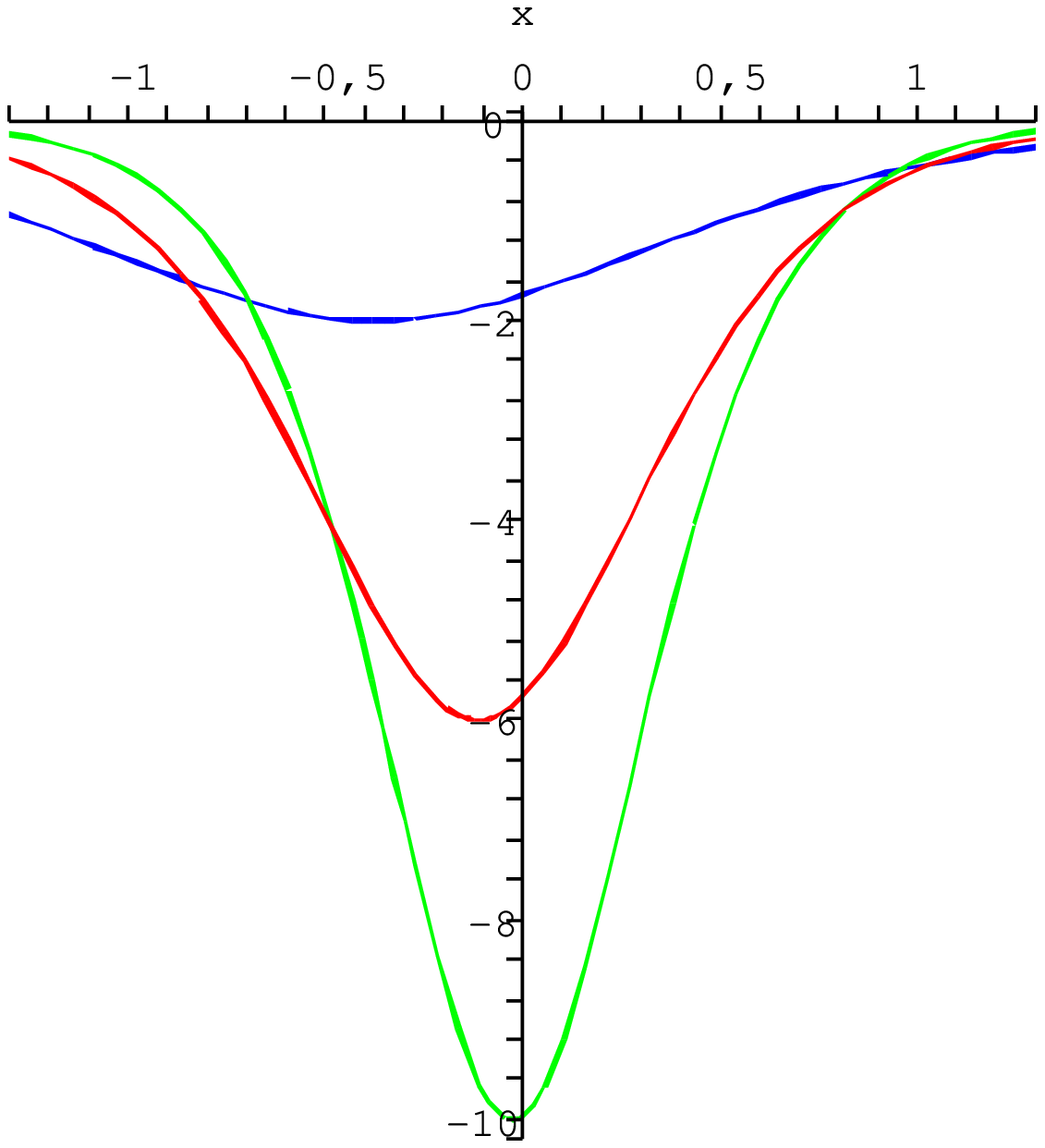}}
\caption{\small
Reflectionless potential:
(a) --- displacement of the hole with its passing through zero
(the values are used: ${\cal E}^{(1)}= -7, -20, -1000$; ${\cal E}^{(2)}=-5$;  $a = 10 + i 100$, $\Re{(b)} = 3$);
(b) --- continuous pulling down of the hole close zero
(the values are used: ${\cal E}^{(1)}= -20$; ${\cal E}^{(2)} = -1, -3, -5$;  $a = 10 + i 100$, $\Re{(b)} = 3$).
\label{fig.reflectionless.1}}
\end{figure}

The found potential (\ref{eq.example.reflectionless.4}) is \emph{reflectionless}: if to start a plane wave from its left asymptotic region to the right, then it passes along the potential without any reflections, reaching the right asymptotic region. This is followed, for example, from \emph{Rule of construction of new reflectionless potentials on the basis of one given one}, formulated in \cite{Maydanyuk.2005.APNYA} (see p.~447; see some corrections in \cite{Maydanyuk.2005.Surveys_in_HEP}; one can make sure in this also, using analysis in p.~278--280 in \cite{Cooper.1995.PRPLC}), taking into account of \emph{continuity} of both superpotentials on the whole axis. This potential is shown in Fig.~\ref{fig.reflectionless.1}.

Let's consider a well-known reflectionless potential of such form (for example, which is presented by (33) in p.~280 \cite{Cooper.1995.PRPLC}):
\begin{equation}
  V_{\rm refl} = A_{\rm refl}^{2} \cdot
    \biggl\{1 - \displaystyle\frac{2}{\cosh^{2}{(\alpha_{\rm refl} x)}} \biggr\},
\label{eq.example.reflectionless.10}
\end{equation}
where $A_{\rm refl}$ and $\alpha_{\rm refl}$ are real constants (connected by condition: $A_{\rm refl} = -\alpha_{\rm refl}$). This potential belongs to the class of \emph{P\"{o}schl-Teller potentials}
(see papers \cite{Andrianov.1999.IJMP} --- in approaches of nonlinear supersymmetry with complex potentials,
\cite{Plyushchay.2007.AP.v322,Plyushchay.2007.JPA,Plyushchay.2007.PLB,Plyushchay.2001.NPB} --- in the approach of hidden nonlinear supersymmetry, \cite{Levai.2004.CzJP.v54n10,Levai.2004.CzJP.v54n1,Znojil.2000.MG} --- in the approach with PT-symmetry, \cite{Cannata.2007.PLA} --- a generalization of Morse potential in two-dimensional space).
One can see that in both asymptotic regions it tends to nonzero limit $A^{2} \ne 0$ (direct limit $A_{\rm refl} \to 0$ \underline{does not give} zero asymptotic limits for this potential, with keeping nonzero hole), its depth at $x=0$ is fixed by constant $A$. Our found reflectionless potential has also hole of finite depth, but one can pull down this hole \emph{continuously} or displace it along axis $x$, passing zero; in both asymptotic regions the potential tends to null values. On the other side, if as the starting potential $V_{1}$ to use nonzero constant potential with high $A^{2}$, then on the basis of the double SUSY-transformations one can obtain the reflectionless potential of the form (\ref{eq.example.reflectionless.1}), but with tails in the asymptotic regions as for the potential (\ref{eq.example.reflectionless.10}). So, use of negative energies of transformation with arbitrary values in the double SUSY-transformations allows to generalize the reflectionless potential (\ref{eq.example.reflectionless.10}).

\begin{figure}[htbp]
\centerline{
\includegraphics[width=74mm]{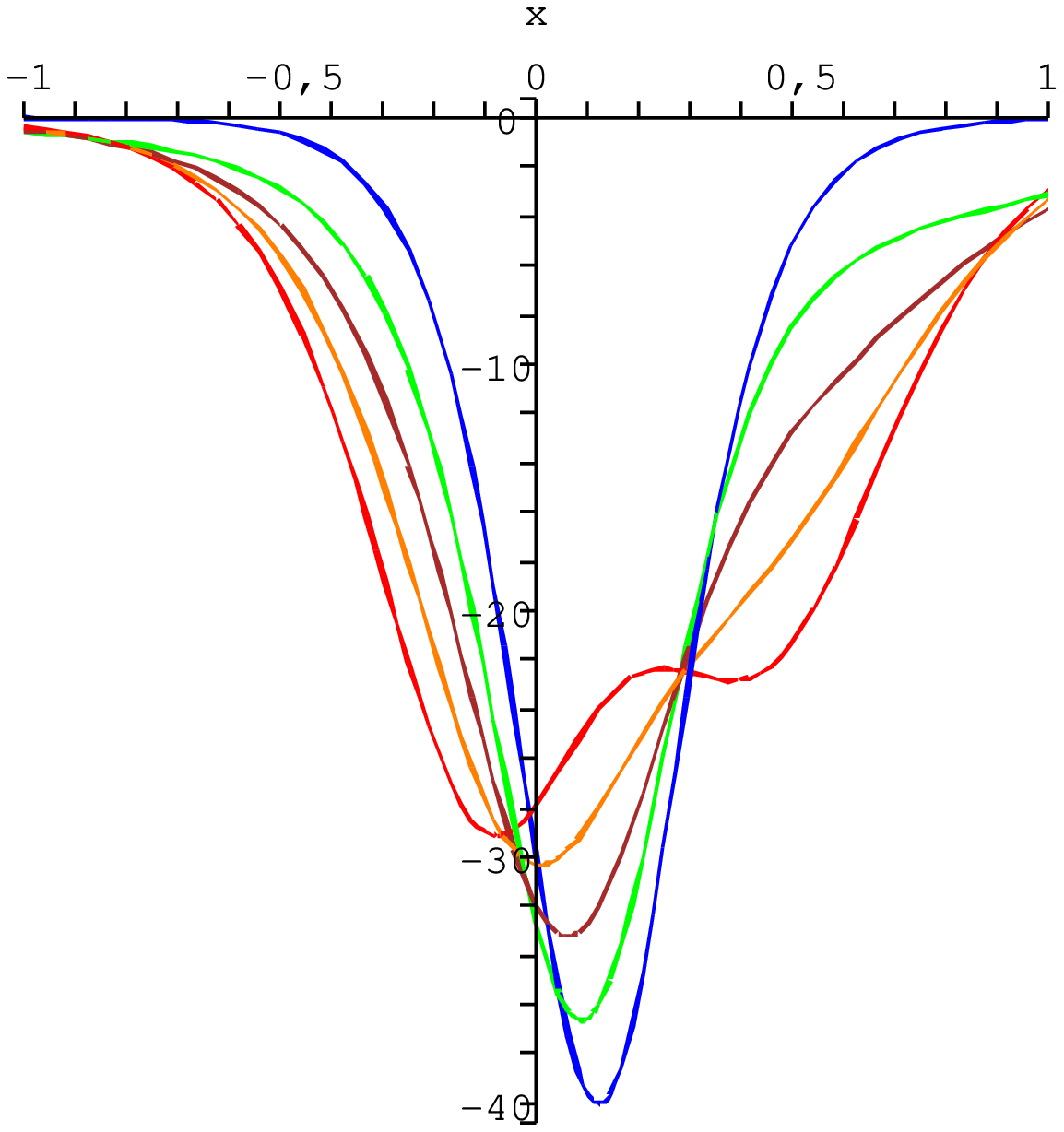}
\includegraphics[width=74mm]{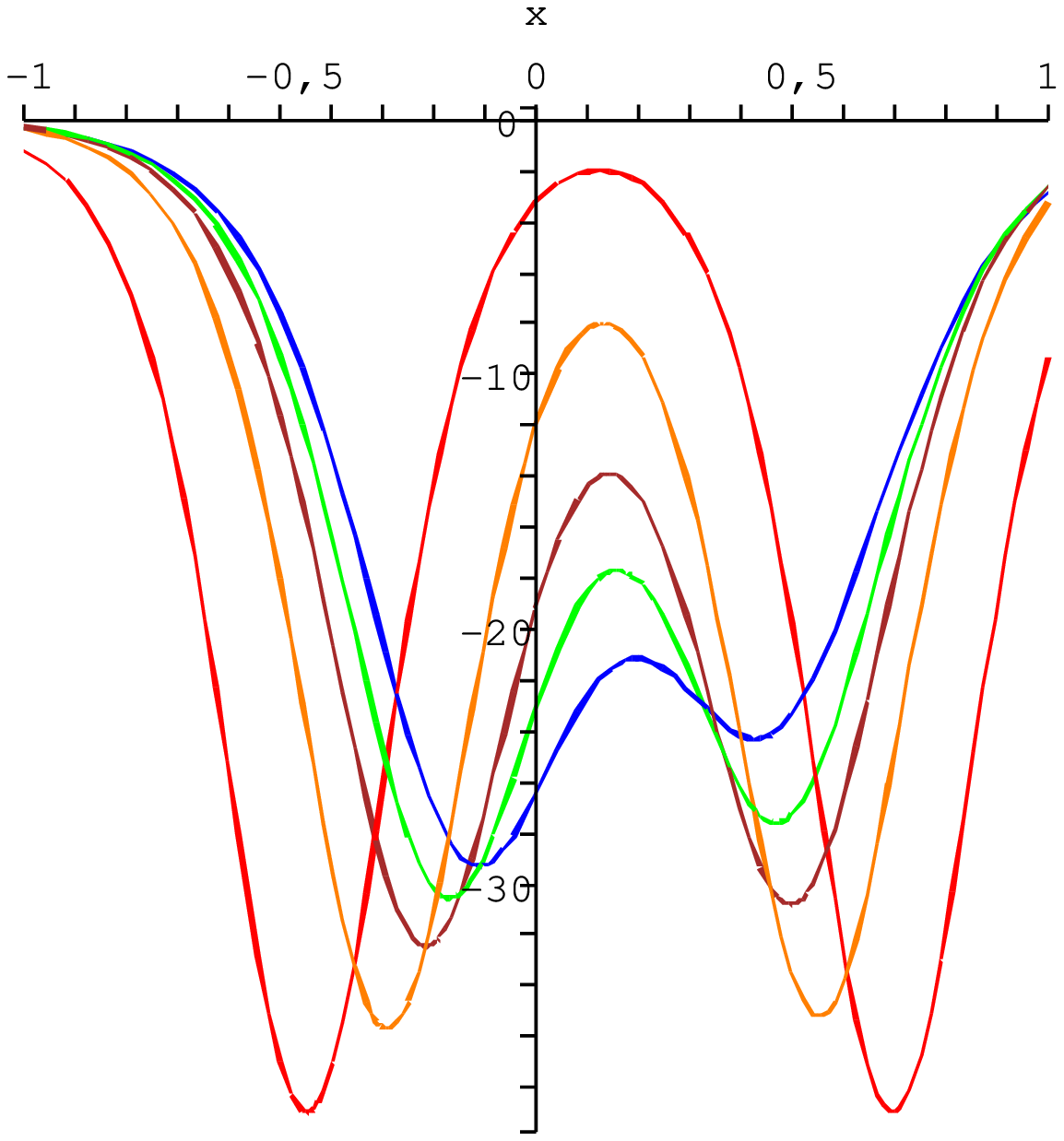}}
\caption{\small
Control of asymmetry of the reflectionless potential
(values are used: ${\cal E}^{(1)}= -20$; $\alpha_{1} = \alpha_{2} = 1$, $\beta_{1} = -\beta_{2} = -3$):
(a) --- continuous increase of the right sloping tail of one symmetric hole with creation of the second small hole in it (${\cal E}^{(2)}= -0.00001, -2, -4, -6, -8$);
(b) --- continuous increase of the second hole with transformation of the whole potential into symmetric double-well one (${\cal E}^{(2)}= -9, -11, -13, -16, -19$).
\label{fig.reflectionless.2}}
\end{figure}

If as the first function of transformation $\phi_{1}$ instead of (\ref{eq.example.1.9}) to use the general its solution:
\begin{equation}
\begin{array}{ccl}
  \phi_{1}(x) & = & \alpha_{1}\,e^{c_{1} x} + \beta_{1}\,e^{-c_{1}x}, \\
  \phi_{2}(x) & = & \alpha_{2}\,e^{c_{2} x} + \beta_{2}\,e^{-c_{2}x},
\end{array}
\label{eq.example.reflectionless.11}
\end{equation}
then we obtain else larger possibilities in deformation of the null potential. We shall show shape of such deformed potential without intermediate calculations. In Fig.~\ref{fig.reflectionless.2} one can see, how by shifting only the second energy of transformation ${\cal E}_{2}$ from zero to the value of the first energy of transformation ${\cal E}_{1}$, one can transform smoothly (continuously) the symmetric one-well potential into symmetric double-well one. So, we have obtained the simple tools for control of asymmetry of the shape of such potentials. Note, that all these potentials satisfy to \emph{Rule of construction of new reflectionless potentials on the basis of the given one} in \cite{Maydanyuk.2005.APNYA} with taking into account of \emph{continuity} of both superpotentials, and therefore they are reflectionless (such potentials look like the reflectionless shape invariant potentials with scaling of parameters \cite{Dutt.1988.AJPIA,Barclay.1991.PHLTA,Barclay.1993.PHRVA,Khare.1993.JPAGB,%
Gomez-Ullate.2004.JPA,Maydanyuk.2005.QTS-5} and self-similar potentials 
\cite{Shabat.1992.INPEE,Spiridonov.1992.PRLTA,Spiridonov.2000.Lectures}
(which can be symmetric in many cases) but here formulas are simpler essentially, and we do not find else in other papers such simple way for control of asymmetry of their shape).

It is interesting to analyze whether a class of the reflectionless potentials, obtained on the basis of group approach developed in \cite{Kerimov.2006.JMP,Kerimov.2007.JPA}, can be expanded after takeing into account of the approach from sec.~\ref{sec.3} for construction of the reflectionless potential (\ref{eq.example.reflectionless.1}).

\section{Conclusions
\label{sec.conclusions}}

In the paper the \emph{approach of double complex SUSY-transformations with not coincident complex energies of transformation} is developed (see sec.~\ref{sec.3}), allowing to deform the given real potential $V_{1}$ (with keeping of its shape real) and its spectral characteristics with obtaining exact solutions. Note the following.
\begin{itemize}

\item
Appropriateness of such approach has been shown with obtaining of new forms of real potentials.

\item
Explicit solutions of the deformation of the shape of the potential, its wave function at arbitrary energy, not coincident with the energies of transformation, wave functions at the energies of transformation are obtained, condition of keeping of continuity of the solutions (i.~e. without appearance of divergencies and discontinuities), isospectral condition with keeping of energy spectra are determined (for potentials with discrete energy spectra).

\item
Using the rectangular well of finite width with infinitely high walls as the starting $V_{1}$ with discrete energy spectrum, by the proposed approach new types of deformation of this potential \emph{with deformation of the energy spectrum as a whole} have been obtained (see sec.~\ref{sec.example.rectangular}; in particularly, we do not find such deformations in variety of deformations of the well, presented in reviews \cite{Zakhariev.1990.PEPAN,Zakhariev.1994.PEPAN,Zakhariev.1999.PEPAN,Zakhariev.2002.PEPAN}).
The new potential contains the rectangular well as own partial case (with simultaneous transformation of the shape of this new potential, energy spectrum, wave functions of all bound states, wave function at arbitrary energy into corresponding characteristics of the rectangular well at needed choice of the parameters of the deformation).

\item
Using the null potential as the starting $V_{1}$ with continuous energy spectrum, new form of the reflectionless real potential has been constructed (see sec.~\ref{sec.example.reflectionless}). This potential generalizes the well-known reflectionless potential of the type $V_{\rm ref}(x) = A^{2}(1-2\,{\rm sech}^{2}{Ax})$, allowing:
{\small
\begin{itemize}

\vspace{-1mm}
\item
to pull down tails of the potential $V_{\rm ref}$ in the asymptotic regions up to zero exactly (with keeping of nonzero depth);

\vspace{-1mm}
\item
to pull down \emph{continuously} the finite depth of the hole;

\vspace{-1mm}
\item
to displace arbitrary along axis $x$ the hole with its passing through zero;

\vspace{-1mm}
\item
to create and to increase the second hole, transforming $V_{\rm ref}$ into double-well potential;

\vspace{-1mm}
\item
to control continuously and simply the asymmetry of the shape of such reflectionless potential.

\end{itemize}}
\vspace{-1mm}
Note relative simplicity of the found potential in a comparison with variety of the reflectionless shape invariant potentials (with scaling of parameters; for example, see~\cite{Cooper.1995.PRPLC}).
\end{itemize}

\section*{Acknowledgements
\label{sec.acknowledgements}}

The author is appreciated to Prof.~Mikhail~S.~Plyushchay for useful comments concerning existence of the hidden nonlinear supersymmetry in the P\"{o}schl-Teller potentials and variety of forms of these potentials.

\bibliography{Darboux-complex-IM_eng}

\end{document}